\documentclass[fleqn,usenatbib]{mnras}
\usepackage{natbib}
\usepackage{amsmath}

\newcommand{\logg}{\ensuremath{\log g}}

\newcommand{\feh}{\ensuremath{\protect\rm [Fe/H] } }

\newcommand{\teff}{T$_{\rm eff}$}

\usepackage{graphicx}
\usepackage{tablefootnote}

\usepackage{footnote}
\usepackage{deluxetable}
\usepackage{longtable}
\usepackage{newtxtext,newtxmath}

\usepackage[T1]{fontenc}
\usepackage{ae,aecompl}

\usepackage{graphicx}	
\usepackage{amsmath}	

\title[]{The S2 Stream: the shreds of a primitive dwarf
  galaxy.\thanks{Based on observations made with the ESO Very Large
    Telescope at the La Silla Paranal Observatory; with Gran
    Telescopio de Canarias at the Observatorio Roque de los
    Muchachos-- Isla de La Palma; and with the Hobby-Eberly Telescope
    at McDonald Observatory.}}

\author[D. Aguado et al.]{David~S. Aguado$^{1}$\thanks{E-mail: daguado@ast.cam.ac.uk},
G. C. Myeong$^{2,1}$,
Vasily Belokurov$^{1}$,
N. Wyn Evans$^{1}$,
\newauthor Sergey E. Koposov$^{3,1,4}$,
 Carlos Allende Prieto$^{5,6}$, Gustavo A. Lanfranchi$^{7}$,  \newauthor Francesca Matteucci$^{8,9,10}$, Matthew Shetrone$^{11}$,
 Luca Sbordone$^{12}$, \newauthor Camila Navarrete$^{12,13}$, Jonay I. Gonz\'alez Hern\'andez$^{5,6}$, Julio Chanam\'e$^{14,13}$,\newauthor
 Luis Peralta de Arriba$^{1}$, 
 Zhen Yuan$^{15}$
\\
\\
$^{1}$Institute of Astronomy, University of Cambridge, Madingley Road, Cambridge CB3 0HA, UK \\
$^{2}$Harvard  Smithsonian  Center  for  Astrophysics,  Cambridge, MA 02138, USA\\
$^{3}$McWilliams Center for Cosmology, Carnegie Mellon University, 5000 Forbes Ave, 15213\\
$^{4}$Institute for Astronomy, University of Edinburgh, Royal Observatory, Blackford Hill, Edinburgh EH9 3HJ, UK\\
$^{5}$Instituto de Astrof\'{\i}sica de Canarias,
              V\'{\i}a L\'actea, 38205 La Laguna, Tenerife, Spain\\
$^{6}$Universidad de La Laguna, Departamento de Astrof\'{\i}sica, 
             38206 La Laguna, Tenerife, Spain \\
$^{7}$N\'ucleo de Astrof\'isica Teo\'rica, Universidade Cidade de S\~{a}o Paulo, R. Galv\~{a}o Bueno 868, Liberdade, 01506-000, S\~{a}o Paulo, SP, Brazil\\
$^{8}$Dipartimento di Fisica, Sezione di Astronomia, Universit\`{a} di Trieste, via G.B. Tiepolo 11, I-34131, Trieste, Italy\\
$^{9}$I.N.A.F. Osservatorio Astronomico di Trieste, via G.B. Tiepolo 11, I-34131, Trieste, Italy\\
$^{10}$I.N.F.N. Sezione di Trieste, via Valerio 2, I-34134 Trieste, Italy\\
$^{11}$University of California Observatories, University of California Santa Cruz, Santa Cruz, CA 95064, USA\\
$^{12}$ESO - European Southern Observatory, Alonso de Cordova 3107, Vitacura, Santiago, Chile\\
$^{13}$Millennium Institute of Astrophysics, Av. Vicu\~{n}a Mackenna 4860, 782-0436 Macul, Santiago, Chile\\
$^{14}$Instituto de Astrof\'isica, Pontificia Universidad Cat\'olica de Chile, Av. Vicu\~{n}a Mackenna 4860, 782-0436 Macul, Santiago, Chile\\
$^{15}$Key Laboratory for Research in Galaxies and Cosmology, Shanghai Astronomical Observatory, 80 Nandan Road, Shanghai 200030, China\\
}
\date{Accepted XXX. Received YYY; in original form ZZZ}

\pubyear{2020}

\begin{document}

\label{firstpage}
\pagerange{\pageref{firstpage}--\pageref{lastpage}}
\maketitle

\begin{abstract}
The S2 stream is a kinematically cold stream that is plunging downwards through the Galactic disc. It may be part of a hotter and more diffuse structure called the Helmi stream. We present a multi-instrument chemical analysis of the stars in the metal-poor S2 stream using both high- and low-resolution
spectroscopy, complemented with a re-analysis of the archival data to give a total sample of 62 S2 members. Our high-resolution program
provides $\alpha$-elements (C, Mg, Si, Ca and Ti), iron-peak elements (V, Cr, Mn, Fe, Ni), n-capture process elements (Sr, Ba) and other elements
such as Li, Na, Al, and Sc for a subsample of S2 objects. We report
coherent abundance patterns over a large metallicity spread
($\sim1$\,dex) confirming that the S2 stream was produced by a disrupted
dwarf galaxy. The combination of S2's $\alpha$-elements displays a mildly decreasing
trend with increasing metallicity which can be tentatively interpreted as a
``knee'' at [Fe/H]$<-2$.  At the low metallicity end, the n-capture elements in S2
may be dominated by r-process production however several stars are Ba-enhanced,
but unusually poor in Sr. Moreover, some of the low-[Fe/H]
stars appear to be carbon-enhanced. We interpret the observed
abundance patterns with the help of chemical evolution models that
demonstrate the need for modest star-formation efficiency and low wind
efficiency confirming that the progenitor of S2 was a primitive dwarf
galaxy.

\end{abstract}




\section{Introduction}

In the current cosmological structure formation theory, dwarf galaxies
are not the main building blocks of bigger systems, although they do
represent a good fraction of the bricks or mortar from which the {\it
  stellar halos of galaxies} like our own are constructed
\citep[e.g.][]{Johnston1996,HelmiHalo1999,Bu05,Bell2008,Co10,Font2011,Belokurov2013,Deason2016}. It
was therefore to great surprise that upon a careful inspection, the
chemical abundance patterns of the surviving high-luminosity (so
called Classical) dwarfs did not match that of the Galactic stellar
halo \citep[see][]{Shetrone2001,she03,venn04,hel06,kir09,To09}. The
arrival of the ultra-faint dwarfs (UFDs) provided a glimmer of hope as
they appeared to contain significantly more metal-poor stars compared
to their more luminous counterparts \citep[see e.g.][]{Kirby2008,simonleo10,Koposov2015,Walker2016}. 
While at low
[Fe/H] a much better match is indeed observed between the UFDs and the
stellar halo in terms of the $\alpha$-elements and carbon 
\citep[see e.g.][]{Sofia2009,freursa10,norrboo10,gilbootes13,salv15,Spite2018,Yoon2018},
the neutron-capture abundances are clearly out of kilter: the bulk of
the UFDs are under-enhanced in Ba and Sr compared to the halo, yet
there are also surprising examples of extreme over-abundance
\citep[see e.g.][]{Her2008,koch13,Roederer2014,jireti16,Roederer2016,Hansen2017,Ji2019}.

Naturally, the question of figuring out the physical conditions at the
epoch of formation for the stars in the halo is best addressed by
studying the actual progenitors, i.e. those unfortunate satellites
that not only were accreted by the Milky Way but were also taken apart
by the Galactic tides at some distant past. As they have breathed
their last, presently these galaxies can only be scrutinized by
identifying their debris scattered around the Galactic halo, piecing
these shreds together and applying numerical models to their
re-assembled cadavers.  Even though this is in fact the main tenet of
the popular sub-field of Astrophysics known as Galactic Archaeology,
not many such forensic studies are currently on record. Large numbers
of detailed chemical abundances have so far only been collected and
analysed for two stellar halo structures: the Sagittarius stream
\citep[see][]{Monaco2007,Chou2007,deBoer2015,Carlin2018,Hayes2020} and
the {\it Gaia} Sausage
\citep[see][]{niss10,Helmi2018,Mackereth2019,Vin2019,mola20,Feui2020}. Additionally,
a small number of stars in globular cluster streams have been under
the magnifying glass of high-resolution spectroscopy
\cite[e.g.][]{roe19}.

This, as one might guess, is not for the lack of trying. In the dark,
pre-{\it Gaia} ages, many a campaign to follow up lower-mass stellar
streams spectroscopically started with high hopes but ended in anger
and disappointment. Some success has been reported for the highest
surface brightness stream -- that emanating from a number of Milky Way globular clusters \citep[e.g.,][]{Odenkirchen2009,Kuzma2015,Ishigaki2016,belokurov2006,sollima2011,mye17b},
yet fluffier sub-structures were not easy to tame
\citep[][]{Casey2013,Casey2014,navarrete15}. The key to success turned out to be
astrometry, as was beautifully demonstrated in the pioneering work of
\citet{helmi99} who used Hipparcos \citep[][]{Hip1997} proper motions
complemented with ground-based measurements to identify a clump of
stars with peculiar angular momenta passing not far from the
Sun. Out of 12 possible members, 3 stars had positive vertical
velocity while 9 had negative vertical velocity. 
Although its physical structure was not apparent, 
due to the data being limited to the Solar vicinity, 
simulations such as \citet{He08} suggested the data were consistent with the partially phase-mixed debris of a heavily disrupted satellite. This supported the interpretation put forward in \citet{helmi99} as multiple wraps of a diffuse stellar stream, sometimes called the Helmi stream. 
The two separate velocity groups could be on different wraps of the now
destroyed satellite, one moving downward, one upward, with respect to
the Galactic plane. The fact that this small stellar clump stood out
from the bulk of the Galaxy due to its high velocity allowed a
significantly more robust membership identification \citep[see e.g.][]{Chiba2000,Re2005,Kepley2007}. 
Building upon these analyses and further refining the kinematic selection of the stream members,
\citet{Roederer2010} carried out a high-resolution spectroscopic
follow-up campaign of the potential member stars of this halo substructure identified by
\citet{helmi99}. Theirs is the first and only example of a detailed
chemical reconstruction of a low-mass disrupted sub-system.

The arrival of the unprecedented high-quality astrometry from the {\it
  Gaia} space observatory \citep[][]{gaia2016} has completely changed
the game of searching for the halo sub-structure. The last three years
have witnessed an avalanche of discoveries of disrupted Galactic
satellites in the Solar vicinity
\citep[][]{He17,mye17,belo18,mye18a,mye18b,mye18d,Koppelman2018,mye19,Borsato2020,Ohare2020,yuan20}
and further out from the Sun
\citep[][]{Ibata2018,Malhan2018,Koposov2019,IbataStreams2019,Ibata2019,Grilmair2019}. The
more distant streams are powerful probes of the Galactic gravitational
potential
\citep[][]{Koposov2010,Gibbons2014,Bowden2015,Bovy2016,Bonaca2018,Erkal2019,MalhanGD12019}. The
nearby sub-structures, on the other hand, present a unique opportunity
to understand the physics governing structure formation in the
high-redshift Universe via the chemical analysis of their (bright and
high surface gravity) member stars. An abundance study of the distant
halo stars is of course possible \cite[see e.g.][]{Battaglia2017} but
is always limited to giant stars at fainter apparent magnitudes thus
making such an endeavour much more laborious.

One of the first thorough sweep through the local stellar halo with {\it Gaia} data 
was performed by \citet{mye18a,mye18b}. Before the {\it Gaia} DR2 became available, 
they used proper motions obtained from the combination of the {\it Gaia} DR1 positions and the
recalibrated Sloan Digital Sky Survey (SDSS) astrometry \citep{deason17,deboer18} further
augmented with the SDSS spectroscopy. 
They relied on Main Sequence stars for which a well-tested photometric parallax relation had been
established \citep[][]{Ivezic2008,Williams2017}. The second most
significant substructure detected with 73 likely members in the
velocity space from this SDSS-{\it Gaia} DR1 dataset \citep{mye18b} was 
a cold and metal-poor stellar stream on a nearly polar orbit, 
plunging through the Galactic disc. The stellar stream
was dubbed S2 as the connection between the stream and the stellar
substructure detected by \citet{helmi99} was not immediately apparent as 
the local volumes covered in two studies were different.
On closer examination, the S2 members picked up by \citet{mye18b} can be
associated with the negative $v_z$ portion of the group originally
described by \citet{He08}. As the SDSS-{\it Gaia} provides more extended volume coverage 
of the local halo, \citet{mye18b} could reveal that the substructure has a clear stellar stream morphology
extended along the direction of its streaming motion, supporting the
interpretation of an additional wrap of a stream. They modelled S2
using a library of tidal debris produced by \citet{Amorisco2017}, and
argued that its progenitor, as a dwarf galaxy, had a total mass of $\sim 10^9-10^{10} M_\odot$ and
a stellar mass of $\sim 10^7 M_\odot$ with an infall time $\sim 8$~Gyr ago. 
In a follow-up study, \citet{mye18d} searched for stellar overdensities in the space of
orbital actions with SDSS-{\it Gaia} DR1, and the S2 stream was once again recovered at high
confidence. This time, a group of stars with positive vertical $v_z$
motion (with comparable action variables) were also included as
potential members in addition to the prominent negative $v_z$
clump. However, the positive $v_z$ portion is much smaller in number, and appears more diffuse and
ambiguous as a structure. 
Though this could be 
contamination from the field stars, it could also be the signature of different wraps as it has long been argued \citep[e.g.,][]{helmi99,He08}.
Under this assumption, \citet{koppelman19} studied the property of the stream with \textit{Gaia} DR2 based on a selection box drawn in angular momentum space. Based on the HR diagram, a range of age from 11 to 13~Gyr was 
suggested for the stream. From N-body experiment, a progenitor with a stellar mass of $\sim 10^8 M_\odot$ 
accreted $5-8$~Gyr ago was suggested, with a claim that the stream contributed $\sim 15\%$ of the field 
stars and $\sim 10\%$ of globular clusters of the Milky Way halo. This would make the stream a significant component of the Milky Way halo.

\begin{figure*}
	\includegraphics[width=145mm,trim={ 0.0cm 0 0.cm 0.cm},clip]{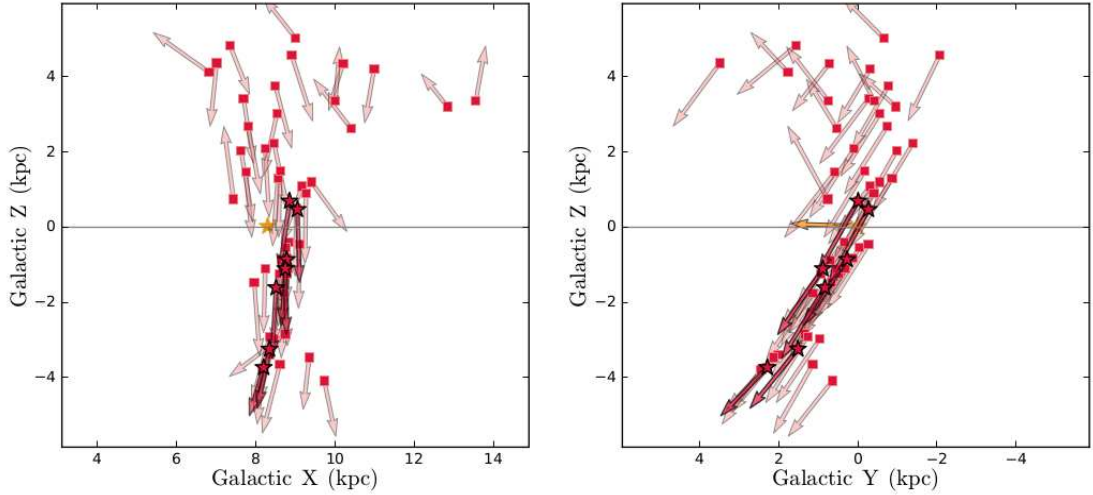}
        \caption{Two-dimensional projections of 52 revised SDSS-{\it Gaia} S2 member stars into the Galactic (X,Z) and (Y,Z) planes. The location of the stars with high-resolution analysis (stars) and SDSS low-resolution analysis (squares) are marked in red. The arrows indicate the individual motions. The Sun and its motion are marked as a yellow star and arrow. The grey line marks the Galactic plane at Galactic Z = 0 kpc.}
    \label{s2spatial}
\end{figure*}
\begin{figure*}
	\includegraphics[width=180mm]{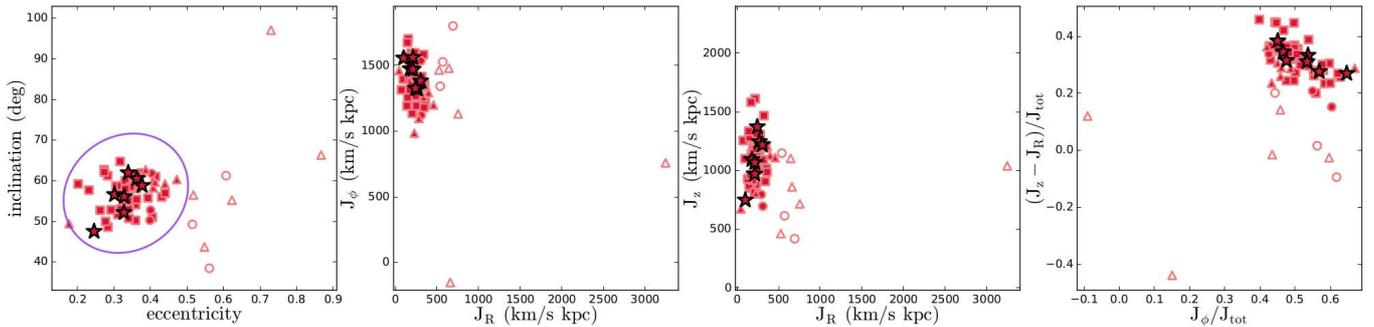}
        \caption{Orbital parameters and the actions of S2 member stars. 52 revised SDSS-{\it Gaia} stars are marked as stars (high-resolution analysis) and squares (SDSS low-resolution analysis). A violet ellipse marks the 4\,$\sigma$ range of a Gaussian model fitted for SDSS-{\it Gaia} S2 stars in the orbital parameter space. 6 APOGEE stars from \citet{yuan20} and 12 stars from \citet{Roederer2010} are marked as circles and triangles. For these 18 stars, the secure members based on the revised kinematics are marked with filled symbols and the rest are marked with unfilled symbols.}
    \label{s2membership}
\end{figure*}
\begin{figure}
	\includegraphics[width=80mm,angle=0]{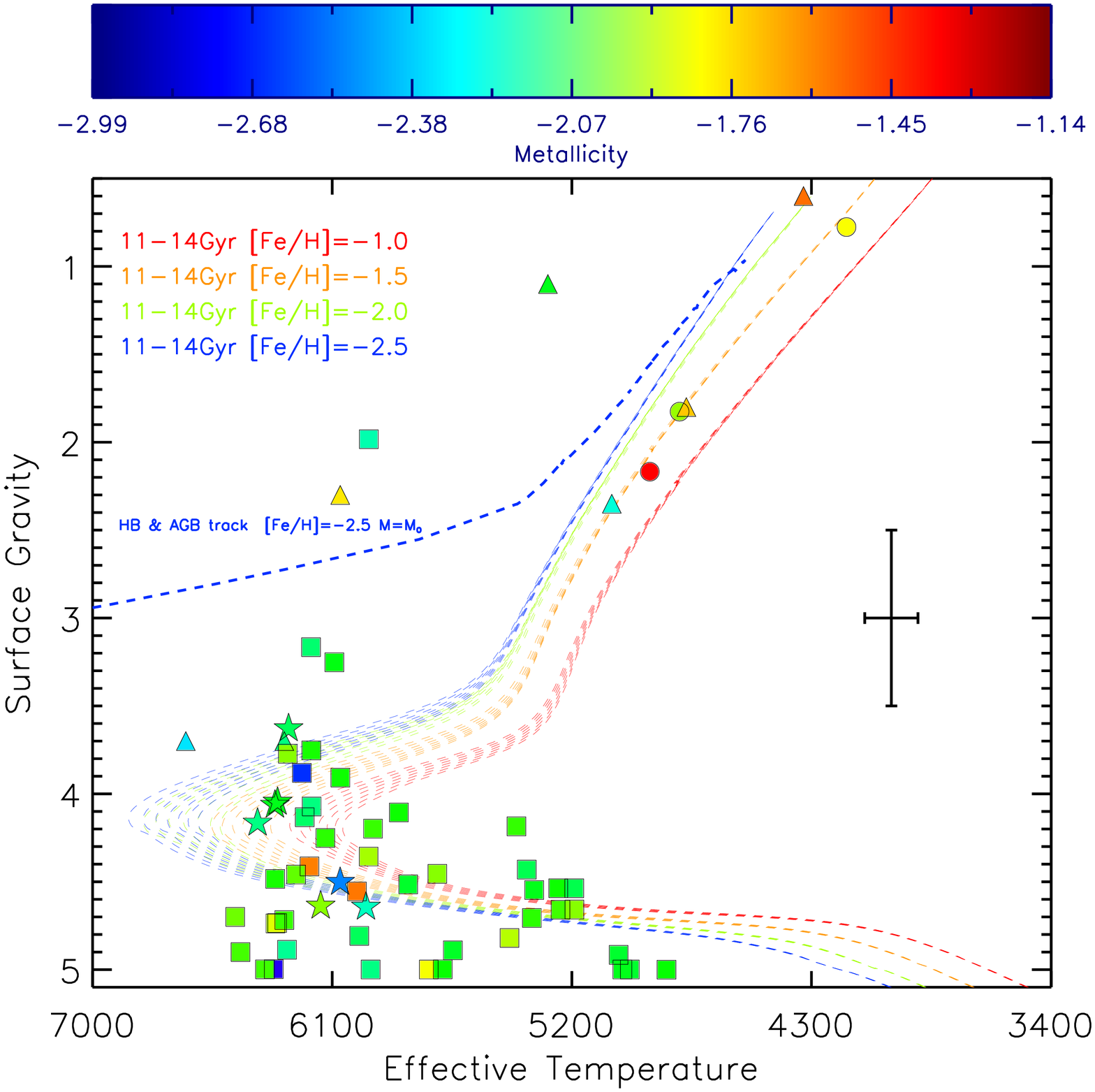}
		\includegraphics[width=80mm,angle=0]{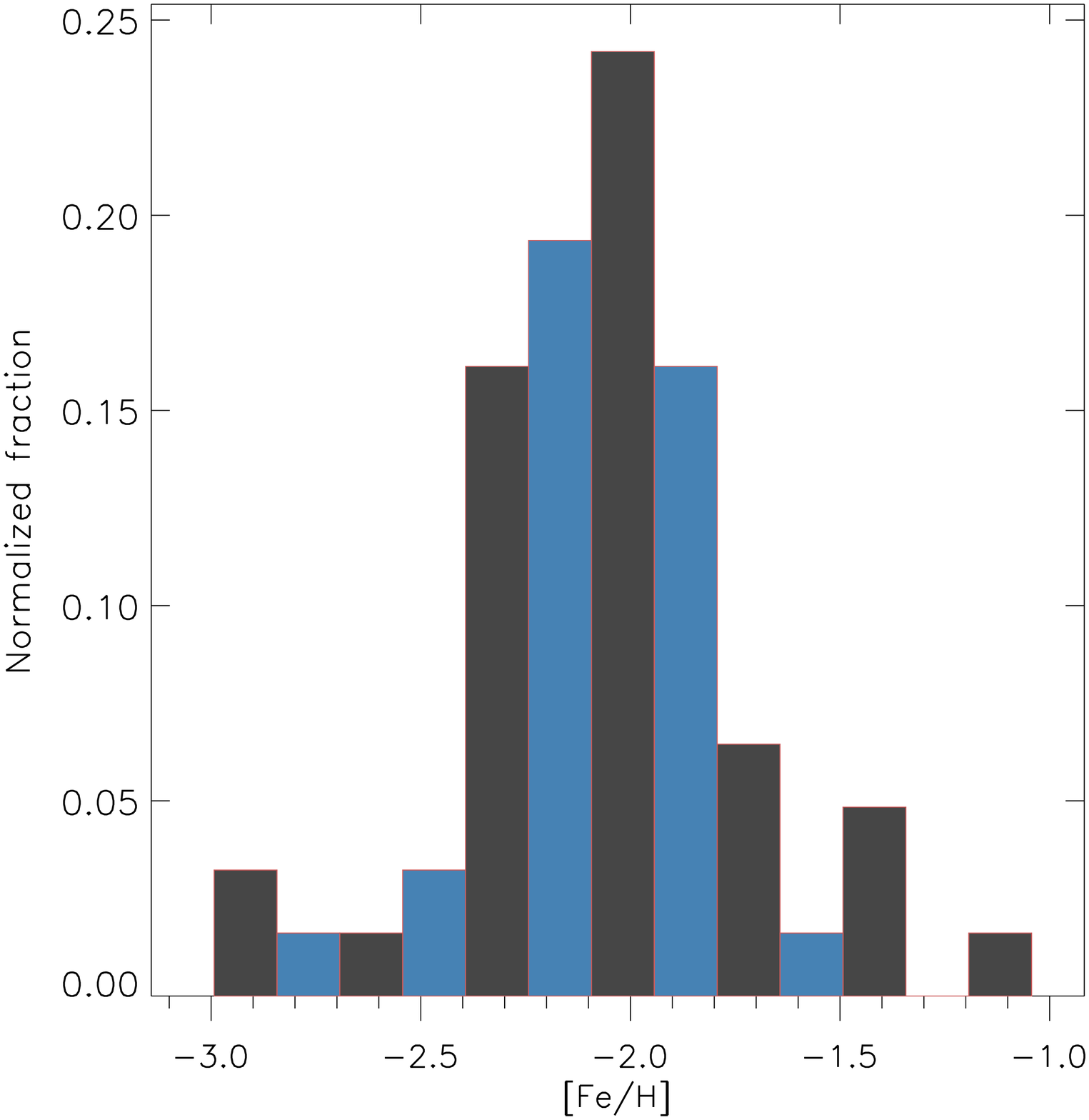}
        \caption{Upper: Kiel diagram, colour-coded by [Fe/H] of S2
          members from SEGUE, BOSS, APOGEE and LAMOST spectra analyzed
          with FERRE. GAIA isochrones for a wide metallicity range and
          different ages from 14 to 11 Gyr (dashed lines) are also
          shown, together with a representative error bars. Lower: The
          metallicity distribution function (MDF) for S2.}
    \label{param}
\end{figure}

The bulk of the high-confidence stellar members in the \citet{mye18b}
sample is within 1-3 kpc from the Sun, the closest to us ultra-faint
dwarf is at a distance of $\sim20$\,kpc \citep{bel07,laev15}, while
the closest satellite similar to S2's progenitor in mass is beyond 50
kpc. In the spectroscopic studies of dwarf galaxies, gaining a factor
of 10-30 in distance can be revolutionizing. This implies that very
bright stars can be measured and therefore even faint traces of
elements can be detected. Moreover, unlike in the distant halo, in the
nearby tidal stream, unevolved stars with higher surface brightness
are available for follow-up at comfortable apparent magnitudes. Dwarf
star spectroscopy circumvents or mitigates the need to correct for
processes that lead to creation, destruction and diffusion of elements
in the stellar interior, just under the atmosphere
\citep[e.g.][]{Spite1982,Thoul1994,busso99,Lind2008,Karakas2010,Michaud2015,Paxton2018}.

Here we report the results of a new spectroscopic campaign to
reconstruct the chemical fingerprint of the progenitor of the S2
stream. We take advantage of the astrometry available as part of the
{\it Gaia} Data Release 2 \citep[GDR2,][]{gaia2018} to revise the
stream membership. Our gold-plated S2 candidate members have been
followed up with high-resolution spectroscopy on 8 and 10-m class
telescopes. We expand the S2's chemical anthology by adding abundances
from low-resolution spectra, both new, obtained by us as well as the
archival data from SEGUE, BOSS, APOGEE and LAMOST re-analysed to
provide a consolidated view of the stream. Additionally, we revisit
the dataset obtained by \citet{Roederer2010} and re-classify their
candidate stream members using the GDR2. As a result our S2 chemical
library covers a wide range of metallicity and contains accurate
abundances of carbon; $\alpha$-elements Mg, Ca, Si, Ti; iron-peak
elements V, Cr, Mn, Co and Ni; as well as n-capture elements Sr
and Ba.

The Paper is organised as follows: in Section \ref{sec:selection}, a
summary of the S2 stream and the target selection for the
spectroscopic follow-up are described. In Section
\ref{sec:observations}, observations and data reduction are
explained. In Section \ref{sec:analisys} we present the two-step
analysis of our S2 catalogue. while a detailed discussion of the
results is presented in Section \ref{sec:discussion} together with the
comparison with chemical evolution models.  Finally, Section
\ref{sec:conclusions} presents the summary and conclusions.

\section{Target Selection}\label{sec:selection}

\subsection{High Quality Sample of S2 Members}
The identification of S2 candidate stars in \citet{mye18b,mye18d} was made from the SDSS-{\it Gaia} catalogue, which is a crossmatch between {\it Gaia} Data Release (DR) 1 \citep{gaia2016}, SDSS DR9 \citep{dr9} and LAMOST DR3 \citep{luo15}. This catalogue of Milky Way Halo stars consists of 61\,911 MSTO and 222 BHB stars with full six-dimensional phase space information and basic stellar parameters \citep[see Section 2.1 of][for more details]{mye18d}. As {\it Gaia} DR2 was not available at that time, the detection made use of proper motions obtained from {\it Gaia} DR1 and SDSS astrometry \citep[see][]{deason17,deboer18}. These stars were reanalysed with later released {\it Gaia} DR2 proper motions, and 52 stars were reconfirmed as high quality members based on their revised action variables \citep{Ohare2020}.

The projected view of the stream in the Galactic coordinates ($X,Y,Z$) is shown in Fig.~\ref{s2spatial}. Here, ($X,Y$) are coordinates in the Galactic plane with $X$ pointing away from the Galactic Centre and $Y$ in the direction of Galactic rotation, whilst $Z$ is perpendicular to the Galactic disc. The majority of the S2 stars form a clear stellar substructure passing through the Galactic disc, just outside of the Solar circle. The velocity vectors of the stars are well aligned with the direction of the stellar stream. The median Galactocentric coordinates of the stream are (X, Y, Z) = (8.7, 0.5, 0.7) kpc, with median actions of ($\mathrm{J_{r}}$, $\mathrm{J_{\phi}}$, $\mathrm{J_{z}}$) = (223, 1394, 1141) kms$^{-1}$kpc, median orbital eccentricity, $e$ = 0.34, and median orbital inclination, $i$ = 58.0 deg.

The main portion of S2 (with negative $v_z$ motion) appears very prominently in the velocity space \citep[see e.g., Fig.~4 of][]{mye18b}. Its distribution is well separated from the main bulk of the field halo stars.
Their spatial distribution in the local halo is also very prominent as a stellar stream which also well coincides with the direction of the streaming motion.
For our high resolution spectroscopic follow-up described in Section~\ref{sec:observations}, we chose targets (marked as stars in Fig.~\ref{s2spatial}) that are well located in the stream to further minimise the potential contamination.

While the majority of S2 stars are forming a prominent stream feature through the Galactic plane, a few stars with comparable action variables are also found outside of the main stream. This could be the result of phase mixing and the trace of different wraps of the stellar stream, as the actions of a star are effectively immune to the phase mixing.\\

\subsection{Literature Survey for S2 Members with Spectroscopy}

We can also use this set of high quality members to identify other S2 stars from the literature with chemical data. In particular, we searched through the 12 S2 members with high resolution spectroscopy claimed by \citet{Roederer2010} and the 6 S2 members with APOGEE spectroscopy found by \citet{yuan20}. 

The kinematics of these stars are revised with {\it Gaia} DR2 astrometry and the distance estimates from \citet{anders2019}. The public software package \textsc{AGAMA} \citep{vasiliev2019} is used for computing the action variables and orbit integration \citep[see,][for more detail]{mye19}. We adopt an axisymmetric gravitational potential model \citep{mcmillan2017} which consists of a bulge, thin, thick and gaseous discs, and an NFW halo as the Milky Way.

Fig.~\ref{s2membership} shows these objects in the planes of action ($J_R, J_\phi, J_z)$, orbital inclination $i$ and eccentricity $e$. The orbital parameters of the 52 SDSS-{\it Gaia} S2 stars are used as a reference to revise the membership of these 18 literature stars. A Gaussian model is fitted for these reference stars in the orbital parameter space. For our study we consider the stars within the 4\,$\sigma$ range of this Gaussian model as the secure S2 member stars.

We retain 10 of these stars as safe S2 members -- 7 from \citet{Roederer2010} and 3 from \citet{yuan20}. They are shown as the filled triangles and circles in Fig.~\ref{s2membership} and listed in Table~\ref{roe}. The remaining stars, shown as unfilled symbols, are not as tightly clustered around the high quality S2 sample.

\begin{table*}
\begin{center}
 \renewcommand{\tabcolsep}{5pt}
\centering
\caption{Stellar parameters of S2 members with high-resolution data from \citet{Roederer2010} and APOGEE.
\label{roe}}
\begin{tabular}{lccccccccccc}
\hline
S2 member&        RA &      DEC &$V$&$T_{\rm eff}$ & $\log g$  & $\left[{\rm Fe/H}\right]$ 
& $\left[{\rm C/Fe}\right]$ &$\rm <S/N>$\\
    &    h  '   ''&$\mathring{}$  '  ''   & mag&K & $\rm cm\, s^{-2}$ &               &                       \\
\hline
         BD$+$30~2611 &15:06:53.8&+30:00:37& 9.1&4330 &  0.60 &   $-$1.45&$-$0.72  &125\\
         CD$-$36~1052 &02:47:37.4&$-$36:06:24&10.0& 6070 &  2.30 &   $-$1.65&$<$0.30  &140 \\
        CS~22876--040 &00:04:52.4&$-$34:13:37& 15.9&6280 &  3.70 &   $-$2.30& $<$0.59  &95\\
        CS~29513--031 &23:25:11.3&$-$39:59:29& 15.1&6650 &  3.70 &   $-$2.55& $<$1.20   &90\\
            HD~119516 &13:43:26.7&15:34:31&9.1& 5290 &  1.10 &   $-$2.15& $<-$0.42   &125\\
            HD~128279 &14:36:48.5&$-$29:06:46& 8.0&5050 &  2.35 &   $-$2.45& $-$0.12   &750\\
            HD~175305 &18:47:06.4&74:43:3&7.2& 4770 &  1.80 &   $-$1.60&$-$0.31  &160\\
            2M02452900-0100541& 02:45:29.0&$-$01:00:54& 10.5&     4168&       0.77&      $-$1.69&$-$0.61&750\\
            2M11503654+5407268&11:50:36.5&54:07:26& 12.1&      4907      &     2.16&      $-$1.14&$-$0.02&265\\
            2M21312112+1307399&21:31:21.1&13:07:39& --&      4975       &      1.82&      $-$1.85&0.10&375\\

\hline
\hline
\end{tabular}
\end{center}
\end{table*}

\begin{table*}
\begin{center}
 \renewcommand{\tabcolsep}{5pt}
\centering
\caption{Coordinates and stellar parameters from low-resolution analysis. This table is an example in which are included stars with further follow-up high-resolution spectroscopy. All the identified S2 members are shown in the appendix in Table \ref{ape1}.
\label{basic}}
\begin{tabular}{lccccccccccc}
\hline
S2 member& & $V$ &      RA &      DEC &$T_{\rm eff}$ & $\log g$  & $\left[{\rm Fe/H}\right]$ 
& $\left[{\rm C/Fe}\right]$ &$\rm <S/N>$&Inst.\\
    & & mag & h  '   ''&$\mathring{}$  '  ''   & K & $\rm cm\, s^{-2}$ &               &         & &             \\
\hline
SDSS  &J0007-0431&       17.5& 00:07:05.36& -04:31:47.7&       6264$\pm$      103&       3.62$\pm$     0.53&      -2.10$\pm$     0.10&      0.71$\pm$     0.27&       55& X-SHOOTER\\
SDSS  &J0026+0037&       15.8& 00:26:19.82&  00:37:34.7&       6305$\pm$      102&       4.04$\pm$     0.52&      -2.02$\pm$     0.10&      0.70$\pm$     0.24&       81& UVES\\
SDSS  &J0049+1533&       15.6& 00:49:39.99&  15:33:17.5&       6315$\pm$      102&       4.06$\pm$     0.52&      -1.94$\pm$     0.11&      0.14$\pm$     0.63&      104& UVES\\
LAMOST&J0206+0435&       15.8& 02:06:53.72&  04:35:44.5&       6143$\pm$      105&       4.63$\pm$     0.51&      -1.99$\pm$     0.11&      0.29$\pm$     0.22&      102& UVES\\
LAMOST&J0808+2418&       15.6& 08:08:33.43& 24:18:46.8 &       6070$\pm$      102&       4.50$\pm$     0.51&      -2.38$\pm$     0.10&      0.20$\pm$     0.35&       71& HORuS\\
SDSS  &J0929+4105&       15.6& 09:29:40.68&  41:05:52.2&       5973$\pm$      105&       4.64$\pm$     0.51&      -2.19$\pm$     0.11&      0.36$\pm$     0.21&      100& HORuS\\
SDSS  &J2345-0003&       17.9& 23:45:52.73& -00:03:05.0&       6380$\pm$      103&       4.16$\pm$     0.53&      -2.31$\pm$     0.10&      0.75$\pm$     0.27&       65&X-SHOOTER \\
\hline
\multicolumn{5}{l}{$\rm <S/N>$ is the average of the S/N of the entire low-resolution spectra.} \\
\multicolumn{5}{l}{Inst. is the instrument used in the high-resolution observations.} \\

\hline
\end{tabular}
\end{center}
\end{table*}

\section{Observations and Data Reduction}
\label{sec:observations}

Our spectroscopic follow-up sample is built from three different sources. The first is Very Large Telescope (VLT) observations with the Ultraviolet and Visual Echelle Spectrograph (UVES) \citep{dek00} and X-SHOOTER \citep{xshooter}. This sample comprises 5 S2 members, albeit with lower signal-to-noise ratio (S/N) than originally anticipated because of weather conditions. We took advantage of the recent start-up of operations of the new High Optical Resolution Spectrograph \citep[HORuS][]{horus} at the Gran Telescopio Canarias (GTC) to obtain data on 2 further objects. Finally, we re-observed 1 S2 object with the second generation Low Resolution Spectrograph \citep[LRS2-B]{LRS2} at the Hobby-Eberly Telescope (HET). The high quality data from LRS2-B allowed us to derive elemental abundances. The 7 stars for which we have acquired data are listed in Table~\ref{basic}.

\subsection{VLT observations}

Three of our targets (J0026$+$0037, J0049$+$1553, and J0206$+$0435) were observed with UVES at the 8.2\,m VLT Kueyen Telescope in three observing blocks (OBs) of one hour in service mode during two different nights, 2019 Aug 2nd and Aug 28th under program ID 0103.B-0398(A). 
A $1\farcs2$ slit was used with $1\times1$ binning in grey sky conditions and an airmass of $\sim1.4$. Our settings used dichroic $\#Dic1\,(390+580)$ and provided a spectral coverage between 330 and 680~{nm}. We corrected each spectrum for the barycentric velocity. The average signal-to-noise per pixel in the spectra was $\sim$15, 30 and 35 at 390, 510 and 670~{nm} respectively. The resolving power for this set up is R$\sim45,000$ for the blue part of the spectrum ($330-452$~{nm}) and R$\sim41,500$ for the red ($480-680$~{nm}). The seeing was $0\farcs95$ during the first and second OB and $1\farcs05$ in the third one. The data were reduced using the REFLEX environment \citep{reflex} within the ESO Common Pipeline Library. 

Two more targets (J0007$-$0431 and J2345$-$0003) were observed with X-SHOOTER during 2019 Aug 29th, and Sep 24/26th under program ID 0103.B-0398(B). Due to the relative faintness of these targets, three OBs of one hour each were allocated but only one per target were finally carried out. The second target has an additional half-time exposure that was aborted due to weather restrictions. We combined both spectra after barycentric corrections. Grey time, a maximum airmass of 1.3 and seeing of $1\farcs$ were the observational conditions. We used slit widths of $0\farcs8$, $0\farcs9$ and $0\farcs9$ (for UVB, VIS and NIR) leading to $R\sim$6200, 7400 and 5400 respectively. The average S/N obtained per pixel in stare mode was 30, 45, 40 at 390, 510 and 670~{nm} respectively. The data were reduced by the ESO staff, and retrieved through the Phase 3 query form.

\subsection{GTC observations}

The observations with HORuS\footnote{http://research.iac.es/proyecto/abundancias/horus/index.php} on the 10.4-m GTC were performed in two different periods. Two 1800\,s exposures of  J0808$+$2418 were taken during the commissioning of the instrument, in 2018 Dec 12th. In addition, four 1800\,s exposures for each of J0808$+$2418 and J0929$+$4105 were taken in queue mode in the period 2019 Nov 18th-Dec 18th, under program 133-GTC130/19B. 

HORuS uses a 3$\times$3 Integral Field Unit (2.1$\times$2.1 arcsec$^2$) with microlenses on both ends of the fiber link that feeds the spectrograph from the Nasmyth focus. The instrument provides a resolving power of $\sim25,000$ and almost continuous coverage between 380 and 690\,nm. We used 8$\times$2 binning to achieve a S/N of 40 at 510\,nm for J0808$+$2418 and J0929$+$4105 in the coadded spectra. Data reduction was done using version X of the {\tt chain}\footnote{Available from github.com/callendeprieto/chain}, which performs bias and flatfield correction, order tracing, extraction, sky substraction, and wavelength calibration.


\subsection{HET observations}
Low-resolution observations were done in queue mode (Shetrone et al. 2007) filling some gaps in the scheduled program devoted to a different project. LRS2-B uses a 12x6 fibre Integral Field Unit and its two arms cover from 370-470 and 460-700\,nm at resolving powers of 1900 and 1100 respectively. J0206$+$0435 was observed from 2019 Jan 16th to 2019 Feb 27th under programmes UT18-3-001 and UT19-1-005. Four spectra of 1800\,s were taken per object with the same setup. 
The LRS2 spectra were reduced with the HET pipeline {\tt panacea}\footnote{https://github.com/grzeimann/Panacea; Zeimann et al. 2020, in prep}. Then we applied by heliocentric correction, coadded and combined the four individual exposures leading to an average S/N of 150 at 390\,nm. 

\section{Analysis}\label{sec:analisys}

\citet{mye18b,mye18d} took the stellar parameters, \teff, \logg\ and \feh from the SDSS and LAMOST automatic pipelines. For very metal-poor (VMP) stars, different levels of carbon-enhancement in the stellar atmosphere affect significantly the opacity of each layer. This effect is not included in the SDSS or LAMOST pipelines. For the sake of consistency, we re-derive \teff\, and \logg\, from the original low-resolution SDSS and LAMOST data (see table \ref{basic}). With these quantities in  hand, we compute elemental abundances from both low- and high-resolution spectra. This is the most suitable way to homogenize -- as far as possible -- the results of this multi-instrument program.

\subsection{Low-resolution analysis}\label{sec:low}
\subsubsection{Stellar parameters}\label{sec:param}

\citet{agu17II,agu17I} provided a way to analyze low-resolution data from metal-poor stars in large spectroscopic surveys. We follow this methodology for the SDSS\footnote{https://dr15.sdss.org/optical/plate/search} and LAMOST\footnote{http://dr4.lamost.org/search} data, downloading the S2 members from the online database. Then we shift the spectra to the restframe using a cross correlation function (CCF) and templates extracted from the grid of synthetic models published by \citet{agu17I}. The overall quality of the spectra is high, except for few objects with clear problems. For example, LAMOST J0808+2418 shows a strong artefact spanning 50\,nm in the blue part. We then mask part of that region prior to analysis. 

Extending the work of \citet{agu17I}, we use a grid of synthetic models with limits: $4500~\rm K <$ \teff $< 7000~\rm K$, $1.0  <$ \logg $< 5.0$, $-4.0  <$ \feh $< +0.5$,  $-1.0  <$ \rm [C/Fe] $< +4.0$. Then, we normalize the spectra using a running mean filter of 30 pixels. The FERRE\footnote{{\tt FERRE} is available from http://github.com/callendeprieto/ferre} code \citep{alle06} is able to derive simultaneously stellar parameters by cubic-Bézier interpolation between the nodes of the grid. We selected the Boender-Timmer-Rinnoy Kan \citep[BTRK,][]{boe82} algorithm to minimize the solution function. An average metallicity is derived in this way together with stellar parameters and [C/Fe]. The analysis is summarized in Table~\ref{basic} (the complete table is available in the Appendix \ref{ape1}). The upper panel of Fig.~\ref{param} shows the Kiel diagram for the high quality S2 sample, colour-coded by [Fe/H]. The lower panel shows the metallicity distribution function with the characteristic sharp peak at [Fe/H]=-2.

\subsubsection{Elemental abundances}

The grid of synthetic models does contain [C/Fe] as a free parameter. With this methodology, we can use all the carbon information in the entire spectra. \citet{agu16,agu19b} provide a prescription to derive carbon abundances in metal-poor stars from low-resolution data, depending on \teff\, and the quality of the spectra. We apply those recipes and derive carbon abundances for more than $75\%$ of the sample with an average uncertainty of 0.2\,dex (see Table \ref{basic}).

We also use the FERRE Spectral Windows \citep[FESWI,][in prep.]{agu20b} to derive Mg, Ca, and Ti when possible. In general, for VMP stars, the magnesium triplet is detectable at S/N higher than 40-50, sometimes even lower. This makes it possible to derive magnesium abundances by just masking the \ion{Mg}{i}b triplet region ($\sim5175$\AA) and re-running FERRE assuming the stellar parameters derived in Section \ref{sec:param}. A visual inspection of the quality of the fit is used to weed out unreliable ones. We provide magnesium abundances for $\sim80\%$ of the sample with an average uncertainty of 0.25\,dex. The same procedure is followed for \ion{Ca}{i} at 4226.7\AA\, and \ion{Ti}{ii} at  3913.4, 4307.8,4443.8, and 4468.4\AA. We derive these species in $60\%$, $22\%$ of the cases, with a medium uncertainty of 0.30 and 0.35\,dex, respectively. Despite the modest accuracy, this provides a valuable collection of elemental abundances to explore the chemistry of S2. In addition, a program of low-resolution but high S/N ($\sim150$) observations was done with the LRS2 spectrograph at HET. The high quality of this data allowed us to measure for an extra object up to 8 elements including C, Mg, Al, Ca, Ti, Fe, Sr and Ba also with moderate accuracy. However, only C, Mg, Ca and Ti abundances from low-resolution data are used for the purpose of this work.

  \begin{figure}
	\includegraphics[width=75mm]{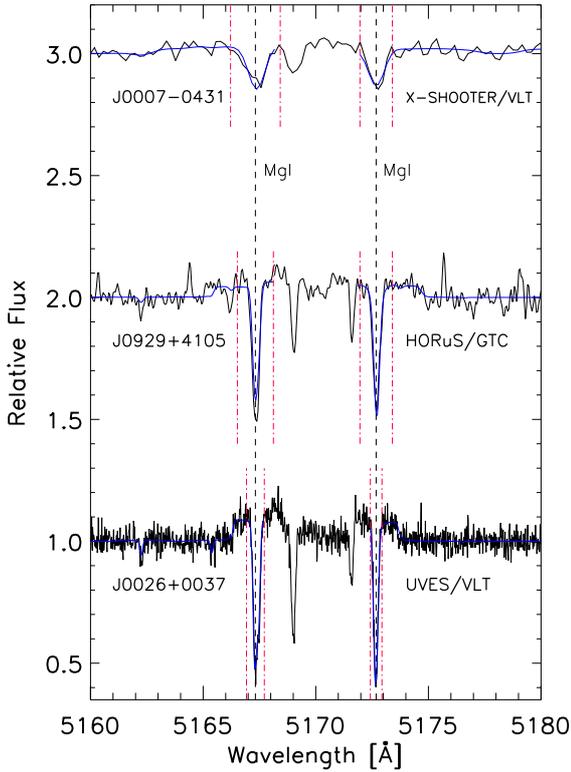}
     \caption{A narrow region of the X-SHOOTER, HORuS and UVES spectra of J0007$-$0431, J0929+4106, and J0026+0037 respectively around the \ion{Mg}{Ib} area (black lines) together with the best FERRE fit (blue lines) only calculated over the spectral windows delimited by red dashed lines. The normalization was performed using a running mean filter.}
    \label{detail}
\end{figure}

  \begin{figure}
	\includegraphics[width=50mm,angle=90,trim={ 0 0 0cm 0.cm}]{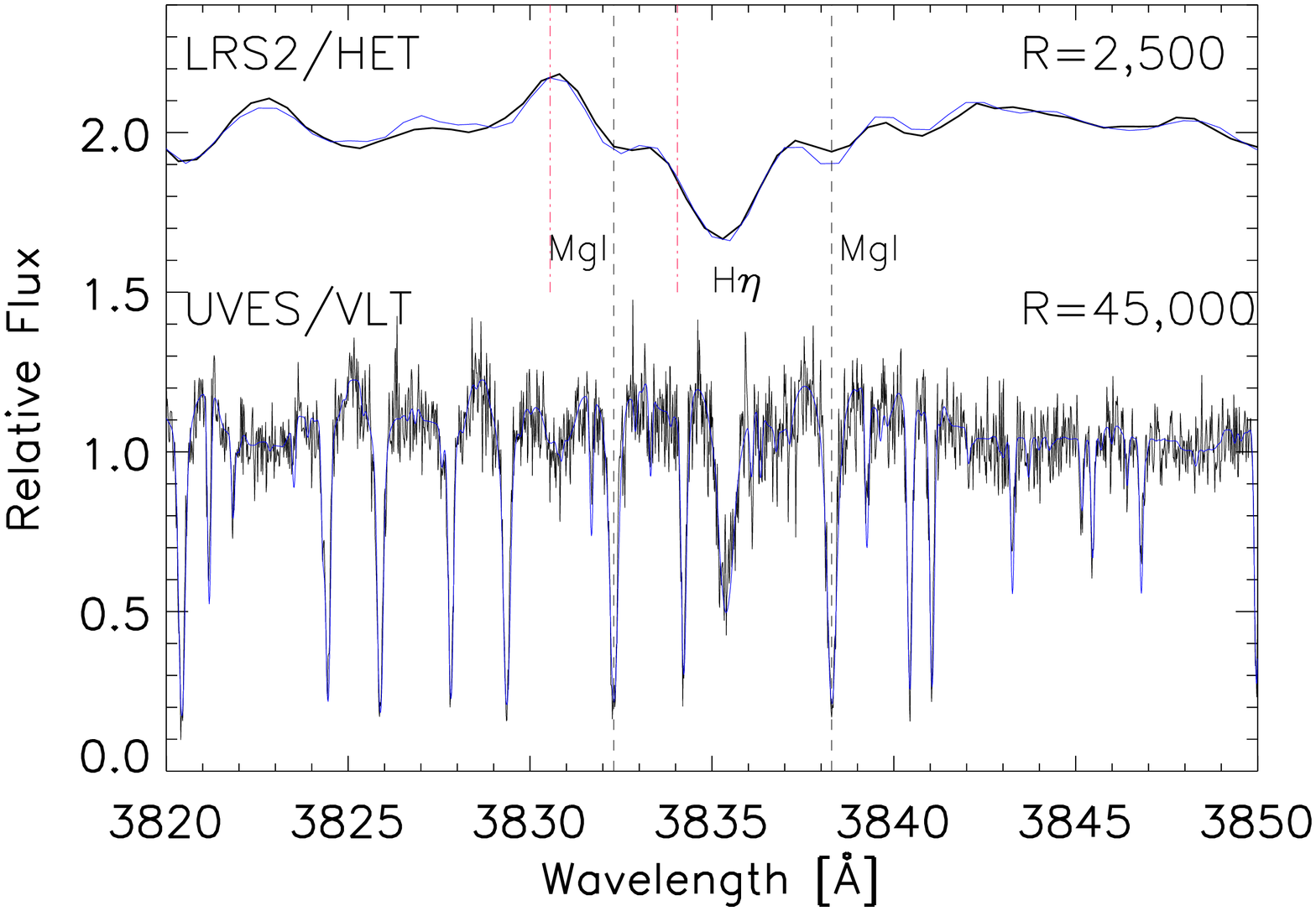}
	\includegraphics[width=50mm,angle=90]{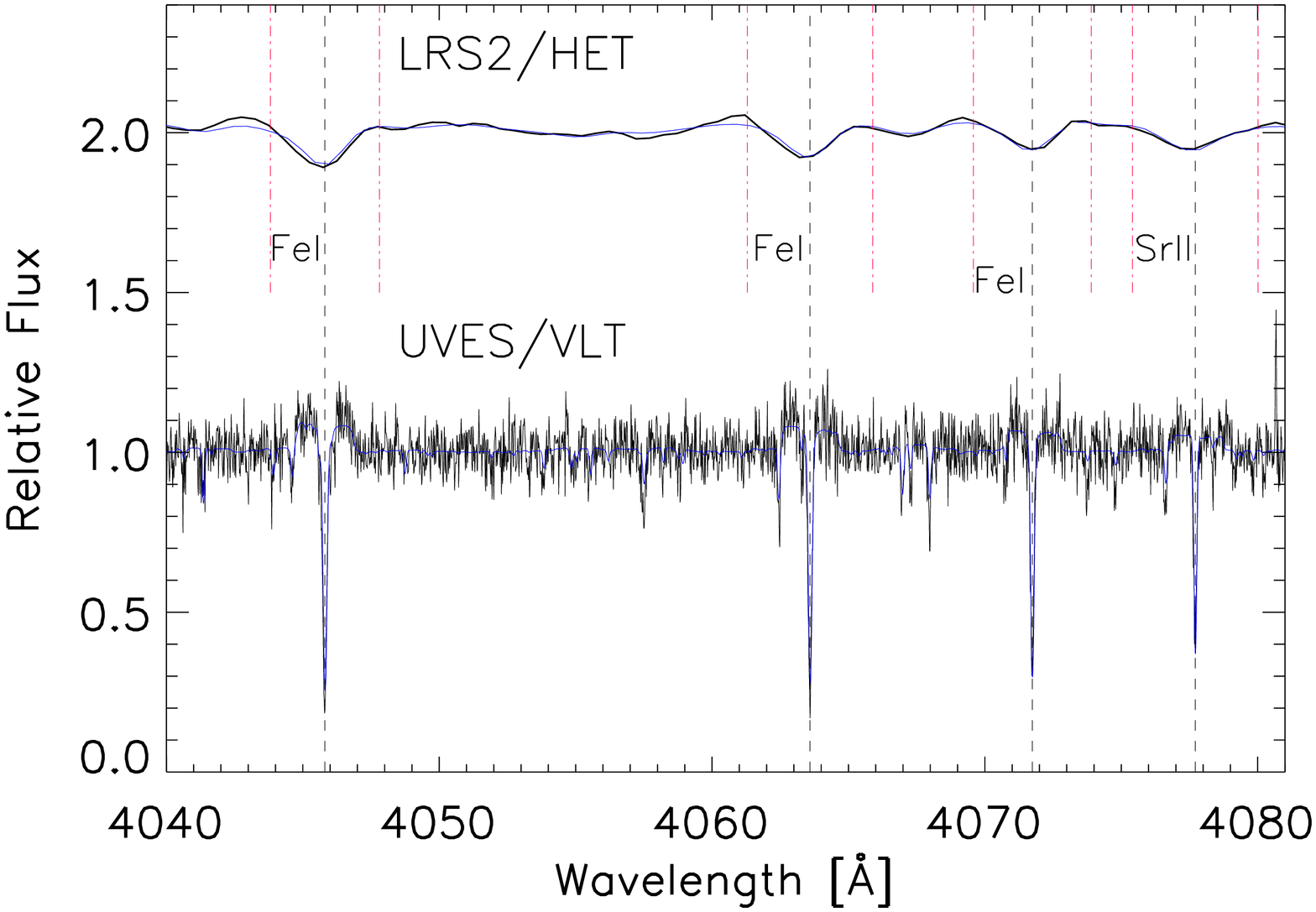}
 	\includegraphics[width=50mm,angle=90]{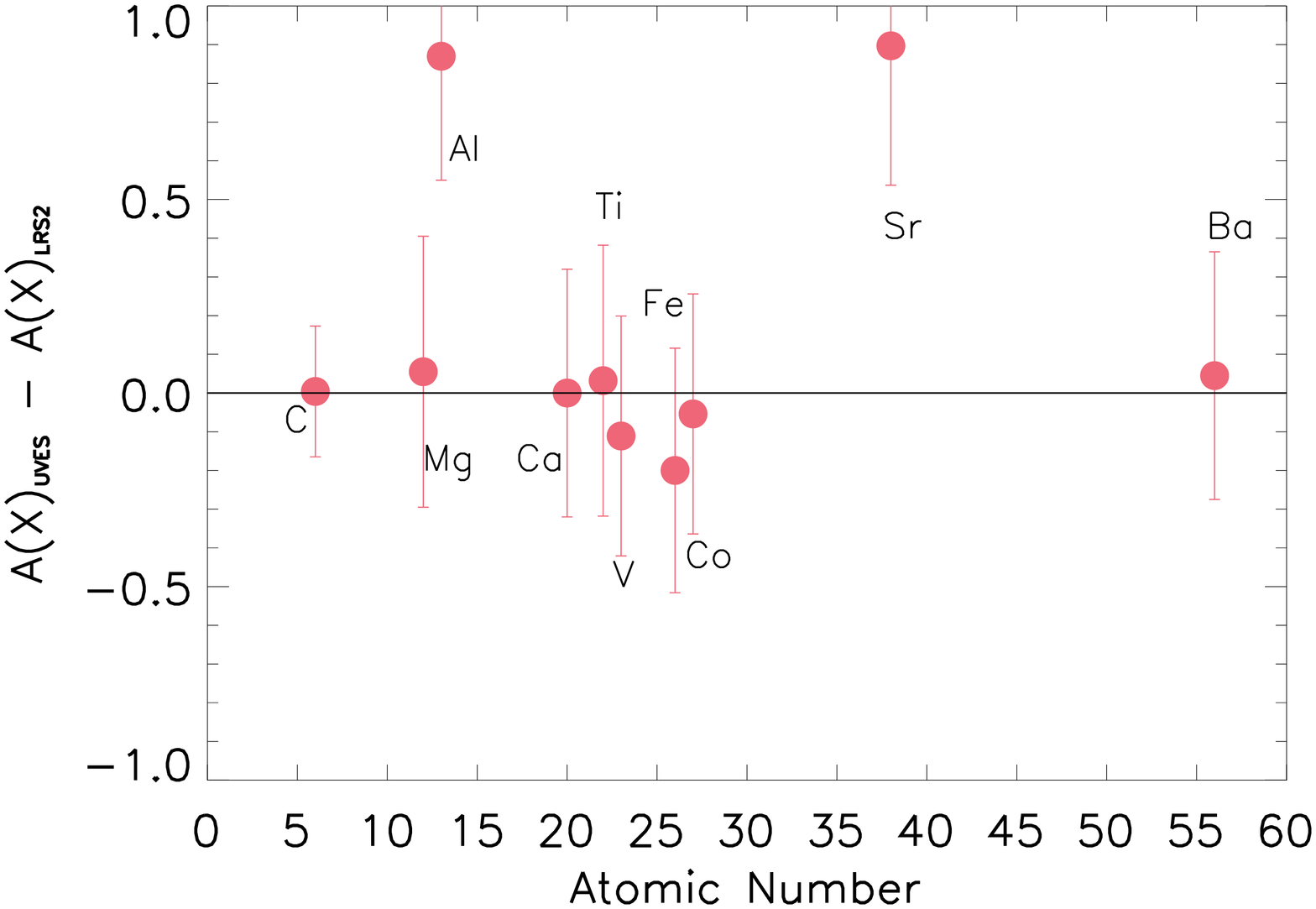}
     \caption{A comparison between the high- and low-resolution (arbitrarily shifted by 1 in the y-axis) analysis performed for J0206+0435. Upper and middle panels show the H$_{\eta}$ region together with \ion{Mg}{I} absorptions and a detail of several \ion{Fe}{I} and \ion{Sr}{II} lines around 4060\,\AA\, respectively. Both UVES and LRS2 spectra are normalized using a running mean filter (black lines) and the best FERRE fit is also shown (blue lines). Red dashed lines delimit the spectral windows where the analysis was performed. The bottom panel shows the difference between both analyses for several elemental abundances. Error bars represent only the low-resolution uncertainties. }
    \label{highvslow}
\end{figure}
 
\subsection{High Resolution Analysis}\label{sec:high}

High resolution X-SHOOTER, HORuS and UVES observations allow us to derive up to 20 different chemical species. The analysis is again carried out with the FERRE Spectral Windows \citep[FESWI,][in prep.]{agu20b}, but applied in a more precise way. We measure chemical abundances of C, Li, Na, Al, Sc, the $\alpha$-elements Mg, Si, Ca and Ti, iron-peak elements V, Cr, Mn, Fe, Co, Ni, and the n-capture elements Sr and Ba. We note that lithium is discussed in Appendix A.
 
We first identify isolated metallic absorptions, discarding some blends that could lead to wrong detections.  Then we build a set of spectral windows, one per target and chemical element. The width of each window is 2.5 times the full width at half maximum (FWHM) centered in the central wavelength of each line. Both the core and the wings of the lines are covered by the spectral windows, leaving the rest of the spectrum masked.  Then, we run FERRE fixing the \teff\, and \logg\, values from Table \ref{basic} to derive carbon and metallicity only within each spectral window. This ensures that we are able to fit all the chemical signals in the data at the same time rather than just performing a line-by-line analysis. The method is robust against possible trends of abundances with wavelength ~\citep[see e.g.,][and references therein]{roe14}. Finally, we can measure lines even in the low S/N end of the data if large enough number of lines cross the spectra, such as for Fe or Ti. 
 
\begin{figure}
	\includegraphics[width=75mm,trim={ 0 0 8.3cm 0.2cm},clip]{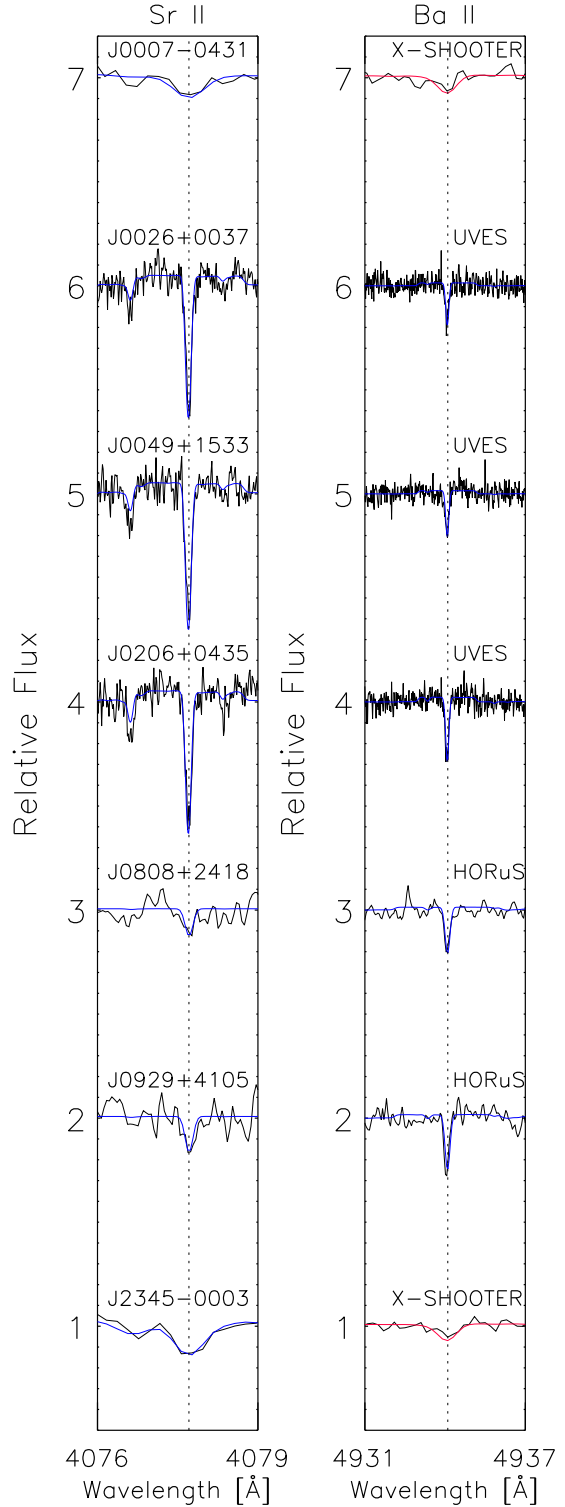}
     \caption{ A narrow region of the X-SHOOTER, HORuS and UVES spectra of the seven stars with high-resolution data respectively around the \ion{Sr}{II} at 4077.7\,\AA\, and \ion{Ba}{II} at 4934.0\,\AA\, areas (black lines) together with the best FERRE fit (blue lines). Upper limits are coloured in red.}
    \label{barium}
\end{figure}

We detect in all cases strong absorption of \ion{Mg}{i} (the triplet around $\sim5170$\,\AA), \ion{Ca}{i} (4226 and 4425\,\AA) and \ion{Ti}{ii} (3759 and 3761\,\AA), and 5 other persistent lines for \ion{Mg}{i}, 7 for \ion{Ca}{i} and 20 for  \ion{Ti}{ii}. In Fig. \ref{detail}, we show an example of two Mg lines detected and fitted within the FERRE spectral windows from data of the three employed instruments. We also report \ion{Si}{i} from UVES and X-SHOOTER data and some lines of \ion{Ti}{i} only from UVES. Carbon abundances are derived in the same way as in Section \ref{sec:low} without the need to use spectral windows but measuring the CH across the entire spectra, especially through the G-band at $4285-4315$\,\AA\ as do the authors in \citet{jon20}. Neither O nor N lines are detected.  The optical spectra contain plenty of iron-peak features  \ion{V}{i}, \ion{Cr}{i}, \ion{Mn}{i}, \ion{Co}{i}, and \ion{Ni}{i}. However, we only detect Co and Ni in the bluer part of the UVES spectra. The sodium absorptions at 5889 and 5895\,\AA\, are clearly detected, even at the moderate S/N of the two X-SHOOTER spectra. We also identify two \ion{Al}{i} absorptions at 3944 and 3961\,\AA. Sr and Ba show the strongest absorptions from any other neutron-capture elements in the optical. We measure \ion{Sr}{ii} at 4077 and, sometimes 4215\,\AA. The most commonly used persistent lines of \ion{Ba}{ii} are 4554, 4930, 5853, 6141, and 6496\,\AA. With HORuS, we cover the two bluer lines, but with UVES, the deepest one at 4554\,\AA\, is out of the spectral coverage. We do not clearly detect Ba in the two X-SHOOTER objects but a robust upper limit is provided from the 4554\,\AA line. Unfortunately, neither Y nor Eu is detected due to the weakness of those features combined with our modest S/N. In Fig. \ref{barium}, we show snippets of n-capture elements (Sr and Ba) lines together with the best fit from FERRE for the three instruments. A complete  list of every line detected in S2 is provided as online material.
 
The elemental abundances in our work are derived assuming one dimensional local thermodynamic equilibrium (1D-LTE). This assumption simplify the modelling of stellar atmospheres and the spectral synthesis of line profiles. It also has the advantage that most of the codes and computed models are public and so easily tested by other authors. An important part of the S2 analysis is based in the determination of $\alpha$-abundances. Mg and Ca NLTE correction tend to be small ($<0.1$\,dex) for FGK stars \citep{mash07, mash13} at this metallicity regime. However, Ti corrections can be up to $\sim0.2$\,dex \citep{sint16}. In addition, the average S/N of our high-resolution data is moderate ($\sim25-30$) or even poor in the bluer parts of the spectra. Nonetheless, taking into account our sample is mainly comprised of VMP dwarf stars with similar \teff\, ($\sim6100$\,K) our accuracy is good enough to detect chemical trends. We also use the online tool INSPECT\footnote{Non-LTE data obtained from the INSPECT database (version 1.0), available at http://inspect.coolstars19.com/}~\citep[see e.g.,][ and references therein]{lind12,amarsi16} to check Li NLTE corrections are also within $<0.1$\,dex. The effect of the 1D-LTE assumption on n-capture process elements in metal-poor stars is not well studied. For Sr and Ba, the corrections could be quite significant \citep{andr11,korot15, gall20} and are probably in the range of $0.02-0.20$\,dex. However, \citet{mash17} derived slightly smaller NLTE corrections for Sr and Ba for a range of metal-poor dwarf galaxy members.
 
 \subsection{High versus low resolution analyses}\label{sec:lowvshigh}
 
To perform a detailed chemical analysis, costly high resolution observations are needed. However, it is possible to work with low-resolution data if we focus in a few elements such as $\alpha$-elements, Fe, Sr and Ba, though the accuracy is moderate. Here, we present a comparative analysis of J0206+0435 observed with both UVES and LRS2. 
 
We detect 370 absorption lines (including 276 \ion{Fe} {I} lines) in the J0206+0435 UVES spectra, while only 43 lines are clearly detected in the LRS2 one (31 of them are \ion{Fe} {I}). We only fit in the spectral windows in which we do identify a line. So, the available metallic information in low resolution data is much more limited due to two main reasons: 1) the much smaller number of detections and 2) the smaller number of pixel per line the spectra contain. Fig.~\ref{highvslow} shows two spectral ranges with several metallic lines, together with the spectral windows (red dashed lines) used during the analysis.  Defining the metallic sensitivity for a certain line as the ability to vary the shape and the FWHM with the abundance of this particular element for a given \teff\, and \logg\, it is clear that there is a large difference between the high- and the low-resolution data.  However, it is also important to note there is substantial information within the LRS2 high S/N data.  In particular, $\alpha$-elements such as Mg, Ca, and Ti and, sometimes, Si can be detected. Finally, the large offset we find in Al and Sr (see Fig. \ref{highvslow}) suggests that we should not rely on these abundances. The most likely explanation for this offset could be related to the bluer region of the spectra where the two Al and Sr absorptions are. We will study this issue in future work.
 
 \subsection{The S2 subsamples from the Literature}
 
 \subsubsection{APOGEE}
 
 Three S2 members selected from \citet{yuan20} have APOGEE DR14 \citep{gunn06, apodr14} information.  These objects were analyzed by the automatic pipeline ASPCAP \citep{aspcap16} that also uses spectral windows to perform elemental abundances analysis from the infrared H-band. We retrieved the data using the online tool\footnote{https://www.sdss.org/dr16/irspec/aspcap} and obtained the stellar parameters together with metallicity and chemical abundances.

 Following the recommendations of APOGEE \citep{apodr14}, we use only calibrated data and discard the titanium abundances as their reliability is uncertain. In addition, we find the fits for Na, V, Cr and Co abundances to be untrustworthy due to the weakness of their absorptions in the NIR. However, for C, Mg, Al, Si, Ca, Mn and Ni, we accept the ASPCAP values.  A summary of the APOGEE S2 members is given in Table~\ref{roe}, while individual abundances are given in Table \ref{ape2}.

\subsubsection{\citet{Roederer2010}} 

 Seven stars already observed and analyzed by \citet{Roederer2010} are S2 members. They used the MIKE spectrograph at the Magellan telescope and provide up to 46 elemental abundances. 
 
 For this paper, we assume their stellar parameters, together with abundance ratios. Two stars, CS~22876--040 and CS~29513--031, are subgiants. However, the other five stars are in a more evolved phase: BD$+$30~2611, CD$-$36~1052, HD~128279, and HD~175305 are on the red giant branch while  HD~119516 seems to be close to the red horizontal branch, at least as judged from Fig. \ref{param}. These stars -- together with those from APOGEE -- help to trace the evolution of S2 as a galaxy. A summary of the stellar parameters associated with these S2 members is listed in Table \ref{roe}, while the elemental abundances can be extracted from Table 7 of \citet{Roederer2010}. 
 
 \subsubsection{Optical versus infrared abundances} \label{optvsnir}
 
 Elemental abundances from both optical and infrared data are considered in this work. \citet{apodr14} and, more recently, \citet{johnson20} studied in detail possible offsets when comparing APOGEE abundances with those from the optical. They found almost perfect agreement in $\alpha$-elements such as Mg and Si. Ca abundances are also recovered at high precision but the accuracy may slightly vary depending on the optical data employed. As discussed in \citet{aspcap16} and successive data releases, both \ion{Ti}{i} and \ion{Ti}{ii} seem to not be reliable in APOGEE so following their recommendation we do not use Ti abundances. Iron-peak elements such as Mn, Fe and Ni behave quite well when compared with optical data sets. On the other hand, V is less precise and is not further used for the purpose of this paper. All of the APOGEE abundances this work contains are zero-point calibrated and therefore admit fair comparison with optical studies.
 
 We have also performed a detailed visual inspection of each ASPCAP fit prior to inclusion in this study. In addition, we have tested both the used stellar models and the fitting algorithm. The experiment consisted of deriving Mg and Ca abundances following the methodology used in this work over the H-band spectrum of 2M02452900-0100541. We find lower abundances at the level of 0.04\,dex in Mg and 0.06\,dex in Ca, fully compatible within the errors. Finally, in the next section we analyse possible NLTE effects in the optical and in the infrared. 
 
 \subsection{Derived uncertainties}
 
 The error bars of the low-resolution analysis shown in Table \ref{basic} are derived by adding in quadrature the internal FERRE uncertainties (statistical error, derived by inverting the matrix curvature) and the possible systematic errors of the stellar parameters of the observed stars. Those systematics are taken as $\Delta_{\rm T_{eff}} =100$\,K, $\Delta_{\rm logg} =0.5$, $\Delta_{\rm [Fe/H]} =0.1$\,dex, and $\Delta_{\rm [C/Fe]} =0.2$\,dex. The rationale for these choices are given in \citet{alle14} and \citet{agu17I}. 
 
 The error bars for the individual chemical abundances from high-resolution data analysis also are derived in a similar way. The typical internal FERRE uncertainty for the UVES data varies between $0.02-0.10$\,dex for iron and neutral titanium respectively. For the HORuS data, the range is $0.07-0.15$\,dex and for X-SHOOTER, $0.04-0.12$\,dex. As we are presenting elemental abundances from different origins, we consider additional sources of systematic errors. First, the 1D-LTE assumption introduces a deviation, as discussed in Sec. \ref{sec:high}. However, the effect on measurements is different in the optical range than in the infrared (H-band). Even if NLTE correction are small, the fact that they could be different in different ranges of the spectrum is an additional source of uncertainty in our comparative analysis. Secondly, we are using different instruments located on different telescopes and this necessarily introduces a hard-to-quantify bias when deriving abundances from different lines. Thirdly, our synthetic models are computed adopting a microturbulence value of $\xi=2.0$\,km/s. The observed microturbulence of dwarf stars tends to be closer to a $\xi=1.5$\,km/s value \citep{cohe04, bar05}. This could lead to an additional uncertainty, estimated to be lower than $0.04$\,dex in metallicity. Combining these possible sources of systematics, we find a systematic error that is at the level of the statistical one $0.03-0.12$\,dex. The errors given in Tables \ref{results1} and \ref{results2} include both the statistical and systematic effects.

\begin{table*}
\begin{center}
\renewcommand{\tabcolsep}{2pt}
\centering
\caption{Abundances, ratios, errors and number of detected lines for individual species derived with UVES.\label{results1}}
\begin{tabular}{ccccccccccccccccccccc}
 &  & \multicolumn{4}{c}{J0026+0037} & & \multicolumn{4}{c} {J0049+1533}  & & \multicolumn{4}{c} {J0206+0435}\\
\cline{3-6} \cline{8-11} \cline{13-16}
Species & $\log\epsilon_{\odot}$\,(X)$^1$ & $\log\epsilon$\,(X) & $\mbox{[X/Fe]}$ & $\sigma$ & $N$ &
                                      & $\log\epsilon$\,(X) & $\mbox{[X/Fe]}$ & $\sigma$ & $N$ & &$\log\epsilon$\,(X) & $\mbox{[X/Fe]}$ & $\sigma$  & $N$ \\
\hline\hline
Li I    & 1.05 &     2.30 &             & 0.10    & 1 &&     2.20 &          & 0.10    & 1           &&   2.26   &         &0.10  &1 &\\
CH      & 8.39 &     6.45 &  $+$0.20    & 0.11    &   &&     6.39 &  $+$0.10 & 0.14    &             &&   6.49   & $-$0.01 &0.11  &  &\\
Na I    & 6.17 &     4.16 &  $+$0.13    & 0.07    & 2 &&     4.19 &  $+$0.17 & 0.06    & 2           &&   4.39   & $+$0.10 &0.09  &2 &\\
Mg I    & 7.53 &     5.77 &  $+$0.38    & 0.05    & 7 &&     5.76 &  $+$0.38 & 0.07    & 7           &&   5.96   & $+$0.31 &0.08  &8 &\\
Al I    & 6.37 &     3.31 &  $-$0.92    & 0.09    & 2 &&     3.36 &  $-$0.86 & 0.08    & 2           &&   3.34   & $-$1.06 &0.09  &2 &\\
Si I    & 7.51 &     5.48 &  $+$0.11    & 0.10    & 1 &&     5.38 &  $+$0.02 & 0.10    & 1           &&   5.60   & $-$0.03 &0.20  &1 &\\
Ca I    & 6.31 &     4.53 &  $+$0.36    & 0.12    & 9 &&     4.52 &  $+$0.36 & 0.09    & 8           &&   4.59   & $+$0.16 &0.09  &9 &\\         
Sc II   & 3.05 &     0.87 &  $-$0.04    & 0.08    & 3 &&     0.84 &  $-$0.06 & 0.15    & 4           &&   0.84   & $-$0.33 &0.13  &3 &\\         
Ti I    & 4.90 &     3.10 &  $+$0.34    & 0.23    & 2 &&     3.09 &  $+$0.34 & 0.21    & 2           &&   3.09   & $+$0.07 &0.25  &5 &\\
Ti II   & 4.90 &     3.01 &  $+$0.25    & 0.08    & 23&&     3.01 &  $+$0.25 & 0.08    & 22          &&   3.21   & $+$0.21 &0.08  &24&\\
V  I    & 4.00 &     1.85 &  $-$0.01    & 0.08    & 6 &&     1.82 &  $-$0.03 & 0.08    & 5           &&   2.00   & $-$0.12 &0.08  &7 &\\
Cr I    & 5.64 &     3.23 &  $-$0.27    & 0.11    & 6 &&     3.17 &  $-$0.32 & 0.12    & 6           &&   3.44   & $-$0.32 &0.10  &6  &\\
Mn I    & 5.39 &     2.68 &  $-$0.57    & 0.08    & 2 &&     2.67 &  $-$0.57 & 0.07    & 2           &&   3.11   & $-$0.40& 0.07  &2  &\\
Fe I    & 7.45 &     5.31 &  $-$2.14$^2$& 0.06    &188&&     5.30 &$-$2.15$^2$&0.06    &187          &&    5.57   &$-$1.88$^2$&0.05&276&\\
Fe II   & 7.45 &     5.16 &  $-$0.15    & 0.09    & 6 &&     5.15 & $-$0.15  & 0.11    & 8           &&  5.33    & $-$0.24 &0.12  &8 &\\
Co I    & 4.92 &     2.83 &  $+$0.05    & 0.12    & 2 &&     --   & --       &  --     &             &&  3.08    &  $+$0.04& 0.12 &2 &\\
Ni I    & 6.23 &     3.84 &  $-$0.25    & 0.13    & 9 &&     3.97 & $-$0.11  &0.11     & 10          &&   3.92   &  $-$0.43&0.12  &11&\\
Sr II   & 2.92 &     0.34 &  $-$0.48    & 0.09    & 2 &&     0.39 & $-$0.42  &0.11     & 2           &&  0.40    &  $-$0.68 &0.10 &2 &\\
Ba II   & 2.17 &     -0.37&  $-$0.40    & 0.10    & 1 &&     -0.37& $-$0.39  &0.11     & 3           &&   0.0    &  $-$0.29 &0.10 &3 &\\
\hline
\multicolumn{5}{l}{$^{1}$Solar abundances adopted from \citet{asp05}} \\
\multicolumn{5}{l}{$^{2}$[Fe/H] from \ion{Fe}{i} is given instead of [X/Fe]} \\
\hline
\end{tabular}
\end{center}
\end{table*}

\begin{table*}
\begin{center}
\renewcommand{\tabcolsep}{2pt}
\centering
\caption{Abundances, ratios, errors and number of detected lines for individual species derived with X-SHOOTER and HORuS.\label{results2}}
\begin{tabular}{ccccccccccccccccccccccccc}
 &  & \multicolumn{4}{c}{J0007$-$0431}  &  & \multicolumn{4}{c}{J0808+2418} & & \multicolumn{4}{c} {J0929+4105}& & \multicolumn{4}{c} {J2345$-$0003}  \\
\cline{3-6} \cline{8-11}\cline{13-16} \cline{18-21}
Species & $\log\epsilon_{\odot}$\,(X)$^1$ & $\log\epsilon$\,(X) & $\mbox{[X/Fe]}$ & $\sigma$ & $N$ & & $\log\epsilon$\,(X) & $\mbox{[X/Fe]}$ & $\sigma$ & $N$ & & $\log\epsilon$\,(X) & $\mbox{[X/Fe]}$ & $\sigma$ & $N$& & $\log\epsilon$\,(X) & $\mbox{[X/Fe]}$ & $\sigma$ & $N$\\
\hline\hline
CH      & 8.39 &     6.88 &  $+$0.74    & 0.11    &   &&5.74 & $+$0.06    & 0.34 &     &&      6.31 &  $+$0.39 & 0.23    &    &&    6.48 &  $+$0.01 & 0.14    &     &\\
Na I    & 6.17 &     3.37 &  $-$0.55    & 0.17    & 2 &&4.42  &$+$0.96    & 0.09 &2    &&      4.42 &  $+$0.66 & 0.08    & 2  &&    4.02 &  $-$0.22 & 0.09    & 2   &\\
Mg I    & 7.53 &     5.65 &  $+$0.37    & 0.13    & 5 &&5.76 & $+$0.38    & 0.07 &7    &&      5.46 &  $+$0.34 & 0.06    & 3  &&    6.03 &  $+$0.42 & 0.12    & 6   &\\
Al I    & 6.37 &     3.42 &  $-$0.70    & 0.15    & 1 &&2.97 & $-$0.69    & 0.12 &1    &&      --   & --       & --      &    &&    3.48 &  $-$0.97 & 0.14    & 1   &\\
Si I    & 7.51 &     5.02 &  $-$0.24    & 0.13    & 1 &&--   &--          & --   &     &&      --   & --       & --      &    &&    5.79 &  $+$0.20 & 0.12    & 1   &\\
Ca I    & 6.31 &     4.41 &  $+$0.35    & 0.11    & 2 &&3.94 & $+$0.34    & 0.14 &2    &&      4.16 &  $+$0.26 & 0.12    & 3  &&    4.51 &  $+$0.12 & 0.09    & 2   &\\         
Sc II   & 3.05 &     1.05 &  $-$0.25    & 0.15    & 1 &&0.49  &$+$0.15    & 0.15 &1    &&      0.39 &  $-$0.25 & 0.15    & 1  &&    --   &  --      & --      &     &\\         
Ti II   & 4.90 &     3.00 &  $+$0.35    & 0.12    & 5 &&2.58 & $+$0.39    & 0.12 &7    &&      2.69 &  $+$0.20 & 0.12    & 4  &&    3.28 &  $+$0.30 & 0.11    & 8   &\\
V  I    & 4.00 &     1.70 &  $-$0.05    & 0.14    & 3 &&1.09   &$-$0.20   & 0.16 &2    &&      1.21 &  $-$0.38 & 0.12    & 2  &&    2.10 &  $+$0.02 & 0.14    & 5   &\\
Cr I    & 5.64 &     3.33 &  $-$0.06    & 0.15    & 6 &&2.83   &$-$0.10   & 0.12 &3    &&      3.14 &  $-$0.09 & 0.11    & 2  &&    3.61 &  $-$0.11 & 0.12    & 6   &\\
Mn I    & 5.39 &     2.68 &  $-$0.57    & 0.08    & 1 &&2.29   &$-$0.39   & 0.13 &2    &&      2.49 &  $-$0.49 & 0.15    & 1  &&    3.29 &  $-$0.43 & 0.07    & 1   &\\
Fe I    & 7.45 &     5.20 &  $-$2.25$^2$& 0.06    & 22&&4.74 & $-$2.71$^2$& 0.13 &59   &&      5.04 &$-$2.41$^2$&0.11    & 66 &&     5.53 &$-$1.92$^2$&0.06    & 40 &\\
Fe II   & 7.45 &     5.29 &  $-$0.09    & 0.12    & 2 &&4.68   &$-$0.06   & 0.15 &3    &&      4.78 & $-$0.26  & 0.15    & 5  &&    5.29 & $-$0.24  & 0.15    & 4   &\\
Sr II   & 2.92 &     0.03 &  $-$0.75    & 0.16    & 1 &&-0.97 & $-$1.18   & 0.19 &1    &&    $-$0.83& $-$1.34  &0.16     & 1  &&    1.03 & $+$0.03  &0.15     & 2   &\\
Ba II   & 2.17 &  $<-$0.08&  $<$0.00    & --      &   &&-0.04 & $+$0.50    &0.13 &2    &&     0.16  & $+$0.40  & 0.14    & 2  && $<$0.25& $<$0.00  & --      &      & \\  
\hline
\multicolumn{5}{l}{$^{1}$Solar abundances adopted from \citet{asp05}} \\
\multicolumn{5}{l}{$^{2}$[Fe/H] from \ion{Fe}{i} is given instead of [X/Fe]} \\
\hline
\end{tabular}
\end{center}
\end{table*}

\section{Discussion}\label{sec:discussion}

\begin{figure*}
	\includegraphics[width=150mm, trim={ 1.5 0.4cm 0.cm 0.cm},clip]{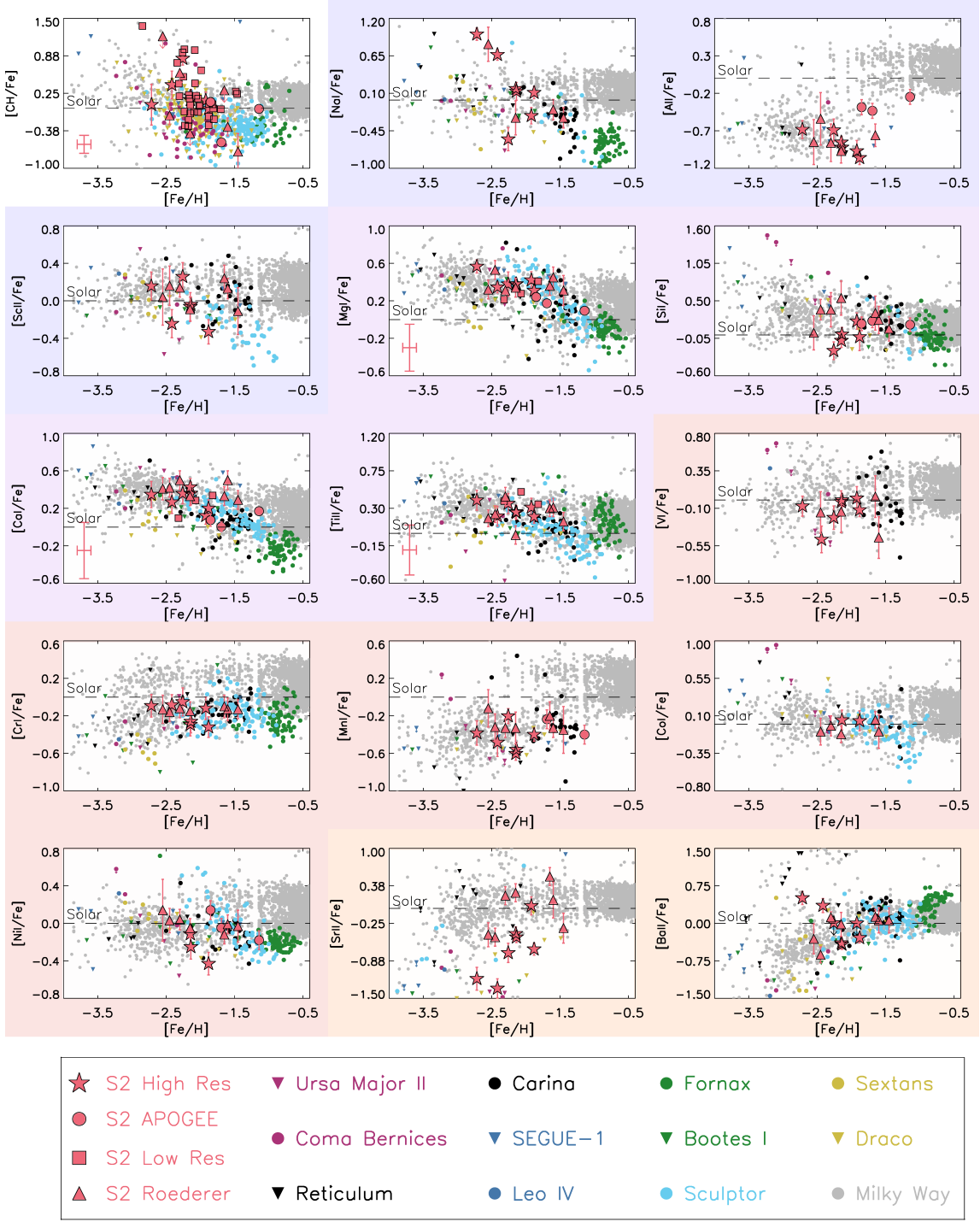}
      \caption{Abundance ratios ([X/Fe]) as a function of metallicity
        ([Fe/H]) for various elements detected in S2 from our
        high-resolution program (red filled stars),
        \citet{Roederer2010} (red filled triangles), APOGEE survey
        (red filled circles) and our SDSS low-resolution
        analysis (red filled squares). The latest are grouped in bins
        of $\sim0.25$\,dex in metallicity and representative error bar
        is shown. We also show elemental abundances from other ultra
        faint and classical dwarf galaxies from literature: Ursa Major
        II \citep{freursa10,kirby15}, Sextans \citep{aokisex09},
        SEGUE-1 \citep{norrbooseg10, fresegue14}, Sculptor
        \citep{geis05, kirsculp12, jabsculp15, kirby15, simonsculp15,skusculp15,hill19}, Coma Bernices
        \citep{freursa10}, Draco \citep{cohendraco09, shedraco13,
          kirby15}, Carina \citep{koch08, venncarina12,norriscarina17}, Bootes
        \citep{norrbootes10,laibootes11}, Leo IV \citep{simonleo10},
        Reticulum II \citep{jireti16}, and Fornax
        \citep{letfornax10,kirby15,letfornax18}. For comparison we
        also show a those of halo
        \citep{aoki07,boni09,aoki12,boni12,cohe13,aoki13,yong13II,yong13III,roe14,agu16}
        and disc stars are shown \citep{bens14, galahdr2}. The background colours
        group different families: odd-, $\alpha$-, iron-peak and
        n-capture elements. Solar abundance references are shown.}
    \label{chemistry}
\end{figure*}

Figure~\ref{chemistry} gives abundance ratios of the elements available
as part of our study split into several groups. We start with carbon
as a representative of the CNO group. The light elements are split
into two sub-groups, with even and odd atomic number. The odd-z group
is represented by Na, Al and Sc, while the even-z elements are Mg, Si,
Ca and Ti. These are followed by the iron-peak group which includes V,
Cr, Mn, Co and Ni. Finally, the last two panels present the abundance
patterns of the two neutron-capture elements, Sr and Ba.

Remarkably, 54 out of 62 stars in our sample are main
sequence/turn-off stars with $\log g>3.5$, offering a unique view as
to the original chemical composition of the environment from they were
formed.  However, for the other 8 evolved stars, there are possible
mixing effects that may alter their chemical signature of lighter
elements, such as Li, C or Na. Surface composition of an FGK star is
considered an accurate guide to the chemistry of the gas from which
they were formed. Once the star leaves the main sequence towards the
red giant branch, dredge-up process cause changes in the convective
envelope. Variations in CNO abundances are firstly small, but could be
altered significantly in more evolved evolutionary phases
\citep{beck80,spi05}. For instance, \citet{pla14b} studied the carbon
abundance problem in giants and provided models to estimate the amount
of carbon converted in nitrogen via the CN cycle \citep{charbonnel95}.

\subsection{The beginning of star formation: to knee or not to knee?}

The star-formation efficiency at the beginning of dwarf's life can be
gleaned from the evolution of the abundance ratio of $\alpha$-elements
with even atomic number to those in the iron peak group. Elements in
both groups can be synthesized either in the interiors of massive
stars or in the explosions of lower-mass stars in supernovae of type
Ia. However, massive star explosions (core-collapse SN) are less
efficient in delivering the iron peak material to the inter-stellar
medium compared to SNe Ia as some of the stellar core collapses to
form the dark remnant. Given that massive stars live much shorter
lives than the SNe Ia progenitors, the ISM enrichment in the first
$\sim$1 Gyr of the galaxy's life is driven by the massive star
nucleosynthesis dispersed by core-collapse SN (CCSN). Accordingly, the oldest stellar
generations are expected to exhibit an enhancement in
$\alpha$-elements compared to iron \citep[e.g.][]{Nomoto2013}. Once
white dwarfs start to explode in supernova of type Ia, copious
amounts of heavier elements are delivered to the ISM \citep[see e.g.][]{Iwamoto1999}, 
thus decreasing the $\alpha$ over-abundance. 
The delay in the onset of SNe Ia explosions compared
to CCSN provides a natural clock: the efficiency with which the galaxy
formed its stars in the first $\sim1$ Gyr of its life can be judged by
the maximal metallicity [Fe/H] before the decrease in
$\alpha$-abundance
\citep[][]{tins79,Matteuci1990,gil91,mcw97,matt01b,Matteucci2003}.

A predicted pattern consisting of a roughly constant over-abundance of
[$\alpha$/Fe]$\sim0.4$ at low [Fe/H] and a steady decline of
[$\alpha$/Fe] to Solar value beyond a characteristic metallicity
achieved by the galaxy before the onset of SNe Ia sometimes called an
``$\alpha$-knee'' has been observed in several Galactic dwarf
satellites, both intact and disrupting
\citep[see][]{venn04,To09,letfornax10,Lapenna2012,Lemasle2012,deBoer2014,Nidever2020}. The
exact shape of the knee depends on the physical conditions governing
the dwarf's early star formation including chemical yields, the initial mass function (IMF),
star-formation rate, as well as temperature, density, inflow and the
outflow rates of the gas \citep[e.g.][]{Matteuci1990,
  mcw97,Kobayashi2006,Lanfranchi2007,Schonrich2009,Revaz2012,Romano2013,Vincenzo2016,Cote2017,Andrews2017,Weinberg2017,Emma2018}.

\begin{figure}
	\includegraphics[width=75mm,angle=90,trim={ 0 1.0cm 0.cm
            4.3cm},clip]{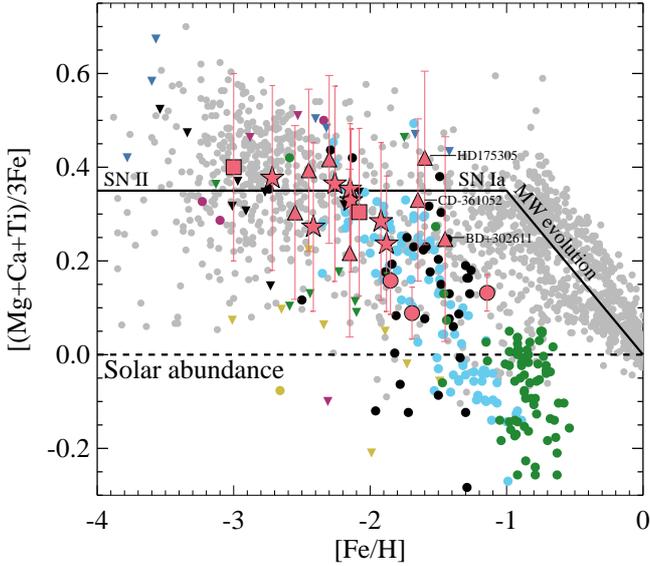}
        \caption{Combined Mg-Ca-Ti $\alpha$-element abundances vs
          metallicity for S2 members. Symbols and references are the
          same as in Fig. \ref{chemistry}. The red square at
          $\rm [Fe/H]=-3.0$ represents the most metal-poor S2 star and the
          y-axis value is calculated by only the \ion{Ca}{ii} K
          line. Red dots from APOGEE data represent Mg-Ca
          abundance. The red square at $\rm [Fe/H]=-2.0$ is a bin
          containing of 10 Mg-Ca-Ti measurements from SDSS
          low-resolution data. The dotted-dashed line indicates the
          solar ratio. The black line represents the canonical Milky
          Way chemical evolution with the "knee" starting at $\rm
          [Fe/H]=-1$ when the effect of SN Ia contribution starts to
          be significant. Due to the lower SFR the S2 knee appears at
          lower metallicities than in the Milky Way. The chemical
          evolution of S2 is slower than the Sculptor due lower
          galactic wind efficiency.}
    \label{alpha}
\end{figure}

As displayed in Figure~\ref{chemistry}, out of four $\alpha$-elements
available, Mg and Ca show some evidence for the existence of a
``knee'' in the S2 stream. Although nucleosynthetic pathways of
hydrostatic (O, Mg) and explosive (Si, Ca and Ti) $\alpha$-elements
are not the same it is common to combine their abundances to improve
the detection quality of the $\alpha$-pattern. Accordingly,
Figure~\ref{alpha} shows a combination ([Mg/Fe]+[Ca/Fe]+[Ti/Fe])/3 as
a function of [Fe/H] for both the S2 members (large red symbols) and
stars from other galaxies (small colored points). The S2 measurement
shown as a square at [Fe/H]$=-3.0$ is the most metal-poor star from
SDSS data where only Ca is clearly detected at [Ca/Fe]$=+0.4$. The
other square symbol at [Fe/H]$=-2.0$ is the average between 9 S2
members with SDSS spectroscopy and with Mg, Ca and Ti
abundances. These stars are within the metallicity range of 0.07\,dex
but with large $\alpha$ uncertainties ($\sim0.30-0.35$) as is shown in
Fig. \ref{chemistry}. The error bar for this representative point is
the dispersion of $\alpha$-values. Finally the filled circles
correspond to the S2 members observed with APOGEE. As explained in Section~\ref{optvsnir},
only Mg and Ca abundances are used for these stars since Ti values are
not reliable  so we compute $\alpha$ as ([Mg/Fe]+[Ca/Fe])/2\footnote{To estimate the impact of that in Fig. \ref{alpha} we perform a simple simulation. We calculated the average deviation of Ti with respect to the mean Mg and Ca abundances from the rest of the sample. The result is a mean deviation of 0.025\,dex. Whether we weight this deviation with the error bars difference is 0.031\,dex. So the fact we do not use APOGEE Ti measurements has very little impact in Fig. \ref{alpha}.}.
Values of $\alpha$ abundance for 99 giant stars in Sculptor obtained with
high-resolution spectroscopy by \citet{hill19} are shown as small cyan circles.
 Additionally measurements for the field stars in the Milky Way are given as grey
dots. The thick black line shows an approximate model of the Milky Way
chemical evolution with a knee starting at the metallicity
[Fe/H]$\sim1$.

Figure~\ref{alpha} demonstrates clearly the presence of the plateau
[$\rm \alpha$/Fe]$\sim0.4$ for S2 stars with metallicities [Fe/H]$<-2$. At higher
metallicities, $-2<$[Fe/H]$\leq-1.5$ (our sample contains only one
candidate member star with a metallicity higher than that), there are
hints of a decline to [$\alpha$/Fe]$\sim0.2$. This can be compared to
the chemical trends observed in the Sculptor dwarf galaxy, which
reaches the $\alpha$ plateau at a very similar metallicity of
[Fe/H]$\sim-2$. In this Figure, the $\alpha$ tracks of S2 and Sculptor
appear aligned, but S2's MDF is clearly truncated around
[Fe/H]$\leq-1.2$ while Sculptor's continues to higher metallicities
and negative [$\alpha$/Fe] ratios \citep[see discussion in
][]{hill19}. It appears that at the outset, the star-formation in the
two galaxies may have proceeded in a similar fashion, but was sharply
truncated in the S2's progenitor as the dwarf was accreted and pulled
apart by the Milky Way.

While Figure~\ref{alpha} points to similarities between Sculptor and
the S2 progenitor in terms of the average [$\alpha$/Fe] evolution, a
careful look at the trends of individual $\alpha$-elements with iron
rouse suspicion. First, [Mg/Fe] is very flat, and save for the two
stars at the extremes of the metallicity distribution, is consistent
with a constant abundance ratio. Looking at Ca, several stars at
higher metallicities appear to exhibit lower values of
[Ca/Fe]. However, the picture is confused by the presence of an equal
number of stars remaining on the plateau at the same
metallicity. Thus, the only conclusion which can be drawn with
certainty is that the [Ca/Fe] spread increases with [Fe/H]. The
clearest difference between S2 and Sculptor is visible in the trend of
[Ti/Fe] with metallicity. Across the [Fe/H] range sampled by the S2
stars, Ti abundance is observed not to be affected by the SN Ia
contribution \citep[as seen in Sculptor][]{hill19}. Note however,
that for the S2 members, [Ti/Fe] remains flat and systematically above
that of the Sculptor stars which also show a downward slope. These
individual abundance trends paint a picture of a dwarf galaxy whose
chemical enrichment was barely affected by the SN Ia pollution. 
\subsubsection{Chemical evolution models}\label{sec:evol}

\begin{figure*}
	\includegraphics[width=160mm,angle=0,trim={ 0cm 12.0cm 0.cm 0cm},clip]{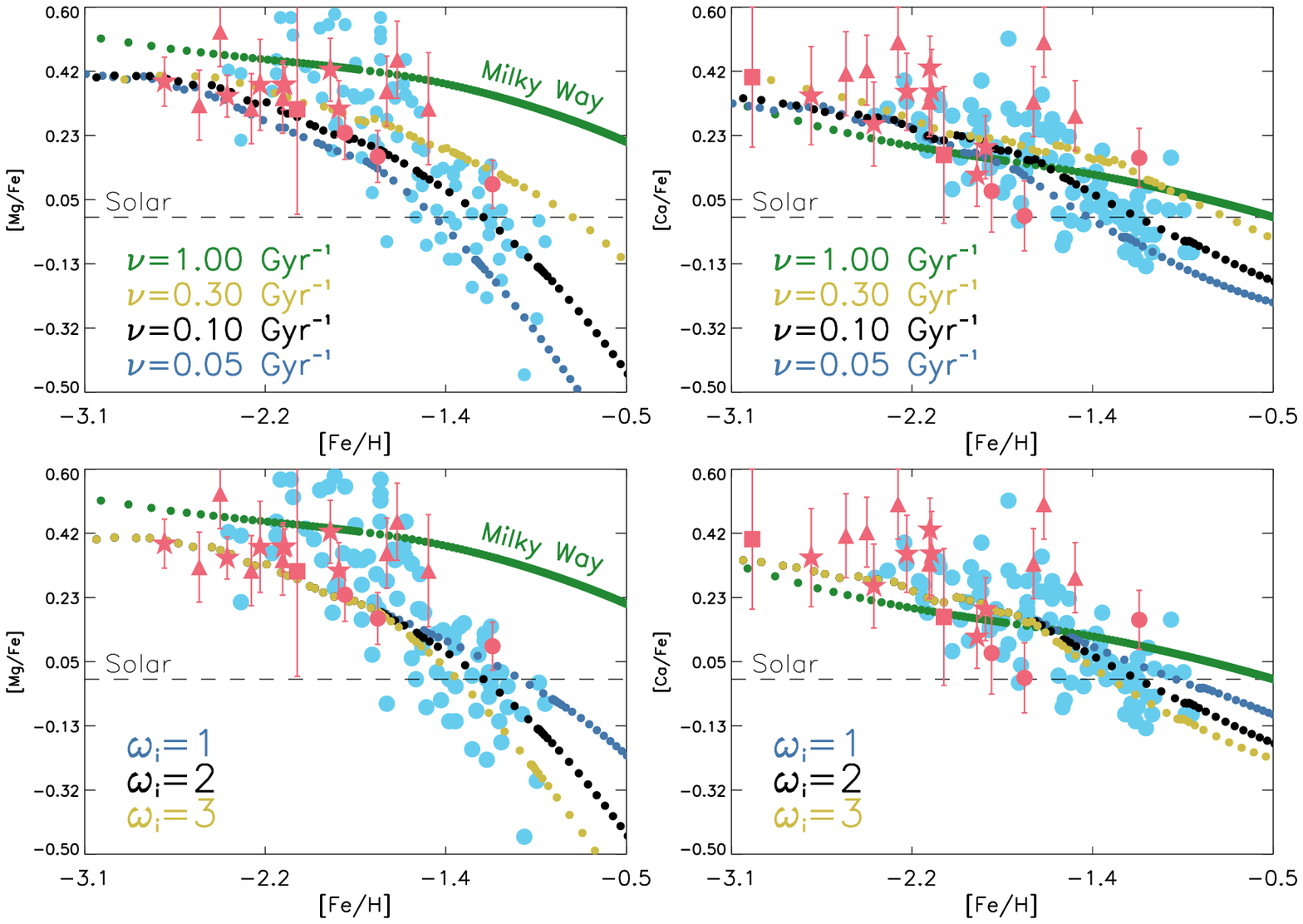}
        \caption{[Mg/Fe] and [Ca/Fe] versus [Fe/H] observed in S2
          compared with the predictions of the chemical evolution
          model by \citet{lan04}. Symbols and references are the same
          as in Fig. \ref{chemistry}. Three different evolution models
          per row are shown. {\it Top:} SFR with $\nu=0.3, 0.1,
          0.05$\,Gyr$^{-1}$. {\it Bottom:} galactic wind efficiency
          with $\omega_{i}=1, 2, 3$. A Milky Way model assuming Salpeter IMF and $\nu=1.0$ is also shown. See \ref{sec:evol} for
          discussion.}
    \label{evolution}
\end{figure*}

Aiming to quantify the chemical evolution of S2, we compare derived
[$\alpha$/Fe] abundance ratios with the results of a detailed chemical
evolution model as presented in \citet{lan04}. As described in
\citet{franc04}, these models take into account stellar lifetimes, the
energy of stellar winds and supernovae types II and Ia, alongside the
nucleosynthetic yields from \citet{thie96,nom97}. By solving the basic
chemical evolution equations given by \citet{tins80} and
\citet{matt96}, the model is able to predict the time evolution of the
individual $\alpha$-element abundances. The main parameters of the
code are the star formation efficiency ($\nu$) and the galactic wind
efficiency ($\omega_{i}$) (i.e. the rate of the gas loss), both
constrained by the data. SFR and wind efficiency are adjusted to fit
the observed values, whereas the initial mass of the galaxy has little
impact in the results of the abundance ratios, but it is indirectly
taken into account through the SFR. A combination of a low star
formation efficiency and a high wind efficiency is required to explain
the abundance ratios of $\alpha$-elements in classical Dwarf
Spheroidal Galaxies, especially the observed lowest values
\citep{lan04}. This is the consequence of the time-delay
interpretation \citep{matt12}, predicting that in the regime of low SFR
the decrease of the [$\alpha$/Fe] ratio is steeper and occurring at
lower metallicity.  Supernovae explosions and winds from massive stars
are important sources of galactic gas outflows, a crucial feature in
the evolution of Dwarf Spheroidal Galaxies \citep{lan03,
  lan07}. Galactic winds occur when the kinetic energy of the gas is
equal or higher than the binding energy of the gas. As a result,
chemically enriched gas and dust are expelled from the galaxy, thus
decreasing the SF rate and changing the evolution of abundance ratios,
in particular [$\alpha$/Fe].

Figure \ref{evolution} shows S2 evolutionary models for Mg and Ca, as well as the best matches for $\nu$ and $\omega_{i}$. The top row shows the
best fits for SF efficiency. We notice that the adopted range of values can account for almost all the data, implying that the SF in the system is compatible with an efficiency $\nu=0.05-0.3$\,Gyr$^{-1}$, with $\nu=0.1$\,Gyr$^{-1}$ as the best value. That means that S2 is characterized by a star formation $\sim$ 10 times lower than that in the Solar neighbourhood~\citep[][]{chia97}. This is shown in Fig. \ref{evolution}, together with a Milky Way model with $\nu=1.0$\,Gyr$^{-1}$ and a Salpeter IMF. However, no galactic wind efficiency has been taken into account for the MW. Spirals are more likely to be affected by galactic fountains with little to no impact on galactic evolution \citep{spitoni13}. Using the same models
\citet{lan04} found that the Sculptor data is best described with a SF
efficiency $\nu_{best}=0.2$\,Gyr$^{-1}$: SF in S2 is similar or
slightly slower compared to Sculptor. The difference in the efficiency
of the galactic wind, on the other hand, is much higher. Whereas in
Sculptor a very high value for the wind efficiency is required, the
bottom row of Figure~\ref{evolution} gives values between
$\omega_{i}=1-3$ (with the best value of 2: almost 6 times lower than
for Sculptor). This difference means that galactic winds in S2 push
away less gas and dust out of the system than in Sculptor, causing
also a lower decrease in the SFR. That fact explains why the
[$\alpha$/Fe]-slope of S2 is less steeper than the Sculptor one shown
in Fig. \ref{alpha}.

\begin{figure}
	\includegraphics[width=57mm,angle=90]{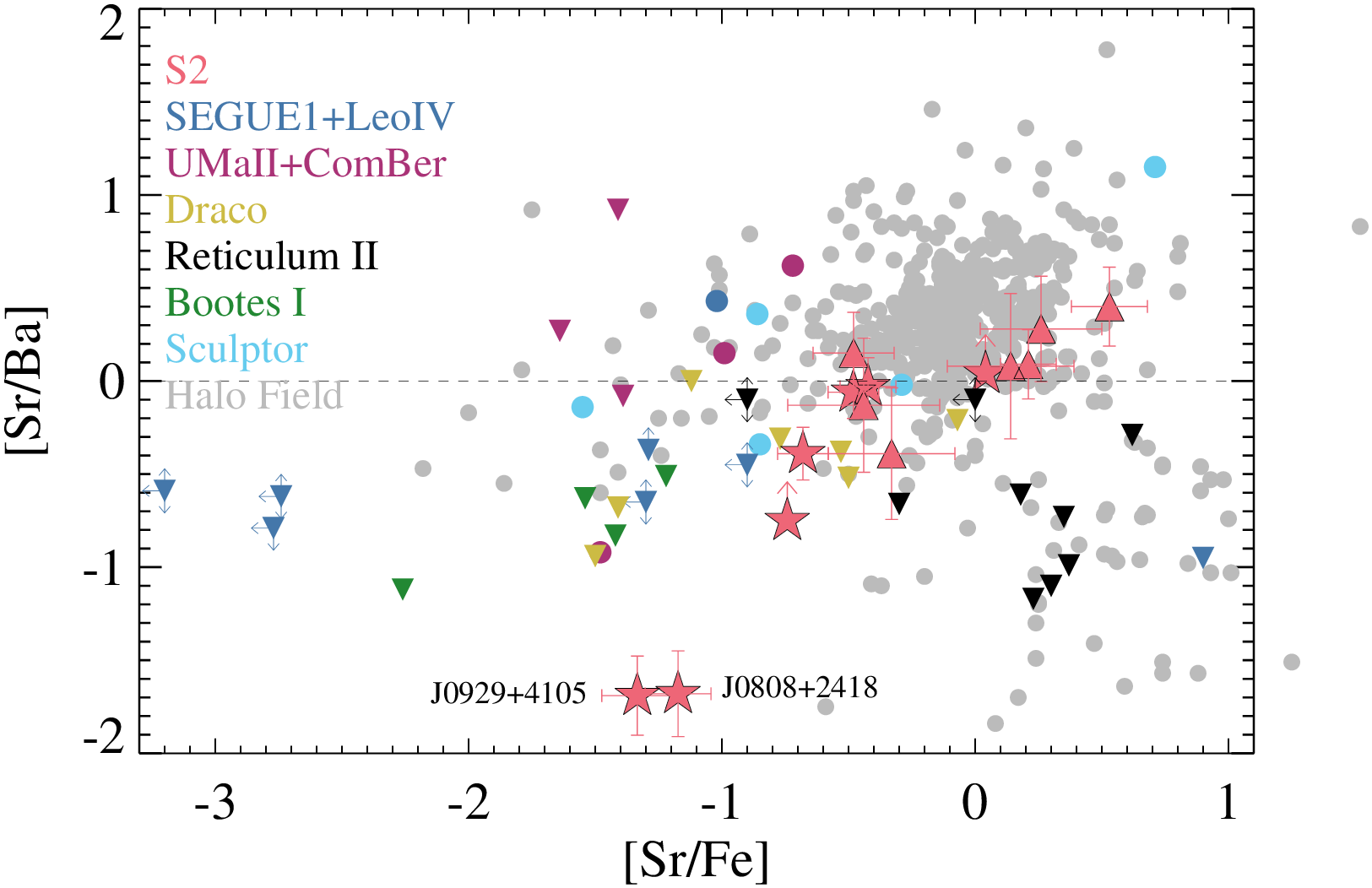}
	\includegraphics[width=57mm,angle=90]{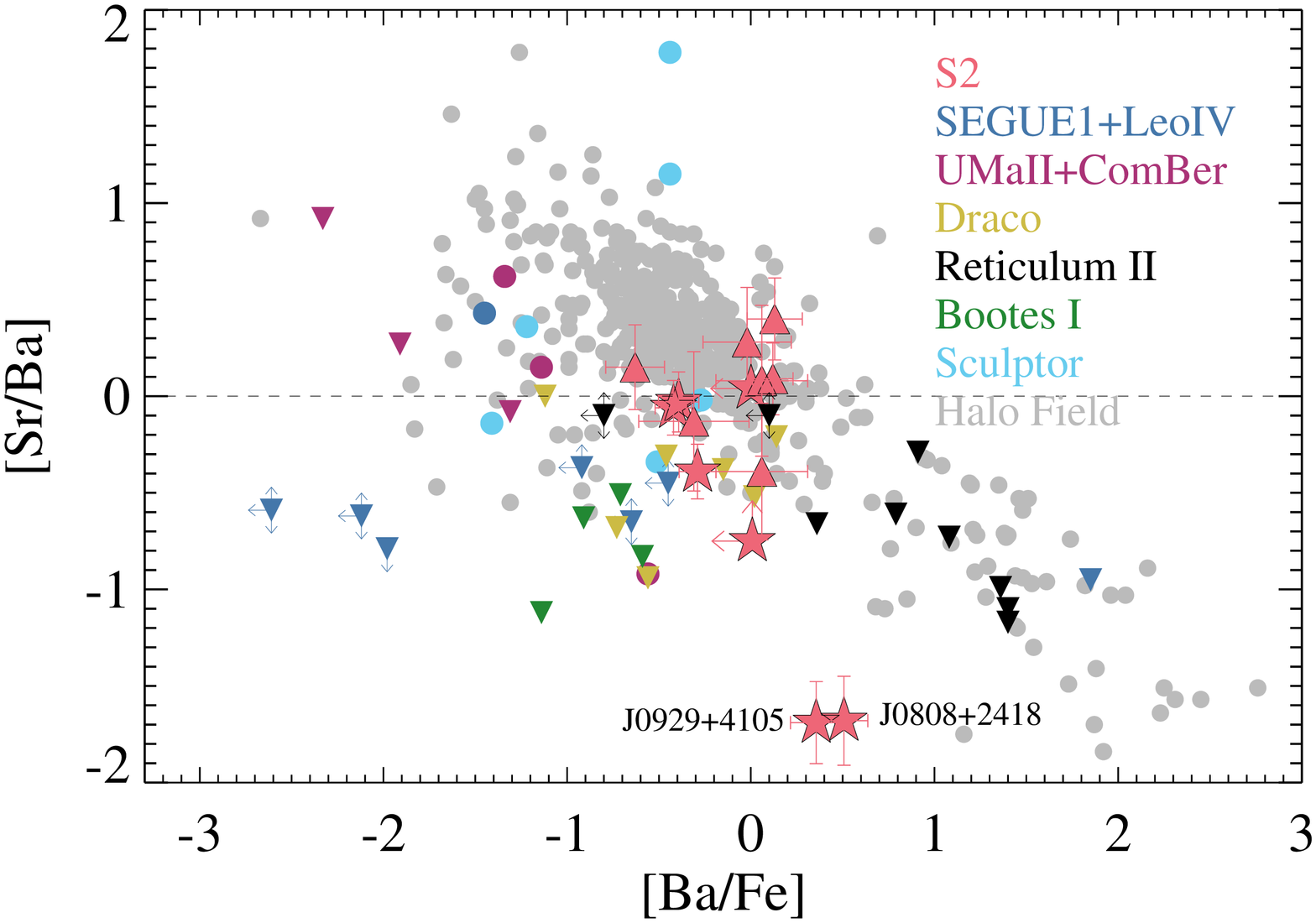}
        \caption{N-capture element ratios for S2 members with
          high-resolution data.  Symbols and references are the same
          as in Fig. \ref{chemistry}. Two S2 members J0808+2418 and
          J0929+4105 show quite unusual n-capture elements pattern with
          extremely low [Sr/Ba] ratio. See Section \ref{sec:neutron}
          for discussion.}
    \label{neutro}
\end{figure}
\subsection{Neutron capture elements}\label{sec:neutron}

The site of n-capture elements production is still under debate ~\citep[see
  e.g.,][and references therein]{tie11}. Evolved AGB stars are
environments in which slow-neutron capture could happen
\citep{busso99}, leading to galactic enrichment of s-elements such as
Sr, Ba and Y \citep{trava04}. The s-process also takes place in the
helium and carbon burning phase of massive AGB stars \citep{kapp11}. On
the other hand, rapid-neutron capture process occurs in more massive
stars via SN II explosions \citep{sne08} or more exotic events such as
neutron star mergers~ \citep{qian00}.  Some fraction of the halo
metal-poor stars show n-capture process enhancement but -- with the remarkable
exception of the Reticulum II galaxy \citep{jireticulum16} -- all the
dwarf galaxies show clear underabundance in n-capture elements
~\citep[e.g.,][]{fresculptor10,simonleo10,norrbooseg10,norrseg10,fresegue14,jiboo16}. There is suggestion that the main
channel for n-capture elements production is core-collapse supernovae
\citep{qian00,arg04}, but in less massive dwarf galaxies, the
production may be dominated by rare events, such as neutron star
mergers \citep{jireticulum16}. Between the rapid and slow flux of
neutrons to form heavier elements is the intermediate-process
\citep{cowan77}. This i-process likely occurs in the most massive AGB
stars with $\sim7-10\,M_\odot$ and could be responsible for n-capture
production ~\citep[see e.g.,][]{jones16}.

In the bottom row of Fig.~\ref{chemistry}, we plot the two n-capture elements, Sr and Ba. We see clear Sr underabundance relative to Fe, well below that of the old halo stars. However, S2 shows higher Ba abundances than other dwarf galaxies. The two most metal-poor stars from S2 with high-resolution data, J0808+2418 and J0929+4105, show a very extreme Sr vs Ba ratio, with [Sr/Ba]$=-1.68$, [Sr/Ba]$=-1.74$ at [Fe/H]$=-2.71$ and [Fe/H]$=-2.41$ respectively.  Fig.~\ref{neutro}
shows [Sr/Ba] versus [Sr/Fe] for both classical and ultra faint dwarf galaxies. The bulk of halo stars show an overabundance of Sr relative to Ba, that is [Sr/Ba]$>0$ \citep{fre10comp,cescu14}.  The spread
of the ratio [Sr/Ba] in Milky Way halo stars may be explained by fast rotating massive metal-poor stars \citep{cescu13} in a nonstandard
n-capture process, but this can not explain ratios with [Sr/Ba]$<-0.5$~\citep[see
  e.g.,][]{fris12}. 
  
However, many dwarf galaxies show even lower
n-capture ratios, as is evident from Fig.~\ref{neutro}. Both Ba and Sr are typical s-process elements with more than 80\% of the Ba in the solar system coming from s-process production \citep{bis14}. But, at very
low-metallicities [Fe/H]$\lesssim-2.8$, there is not expected to be a significant contribution of s-elements from AGB production. Instead, r-process
production in high-energy neutron-rich environments is a dominant channel. Although theoretically such low [Sr/Ba] ratios could be explained by mass
transfer from a binary companion \citep{gallino06} or i-process production in massive AGB \citep{jones16}, we consider both scenarios
less likely since these two stars do not show clear carbon-enhancement
\citep{abate13,jones16} and no hints of radial velocity variation were observed to detect binarity. However, both explanations should not be completely discarded at this point and more observations are required. Finally, \citet{fre15rev} proposed that a special feature
of dwarf galaxies with subsolar [Sr/Ba] ratio might that they reflect the composition of the earliest star-forming clouds. We argue that the n-capture pattern in some of the most metal-poor S2 stars could be due to r-process production and potentially reflect the chemical composition of the pristine gas, polluted only by neutron star mergers or type II supernova. Although the S2 n-capture pattern shown in Fig. \ref{neutro} may suggest exotic Sr-Ba production, we emphasise that we only find two stars in the most interesting part of the diagram.

\subsection{Other elements}

\subsubsection{Carbon}\label{sec:carbon}

The presence of carbon in molecular clouds permits cooling channels allowing low-mass star formation due to fragmentation. \citet{mey06} showed how, in rotating massive stars, internal mixing and stellar winds could be responsible for providing large excesses of CNO to the interstellar medium in early star formation. The fraction of metal-poor stars showing carbon-overabundance increases towards lower metallicities ~\citep[e.g.,][]{coh05, car12,yong13III, boni15}. Different explanations have been proposed, including fallback mechanisms after core-collapse supernovae \citep{ume03} and the existence of a core-collapse sub-type that only ejects the outer layers with lighter elements \citep{ishi14}. There are also extrinsic explanations, such as significant mass-transfer from a more evolved AGB companion~\citep{her05}.  \citet{pla14b} showed that the fraction of carbon-enhanced metal-poor (CEMP) in the halo is 25\% in the VMP regime ([Fe/H]$<-2.0$). According to the classification by \citet{bee05}, CEMP-no stars are those that show no neutron-capture element enrichment. The oldest relics from star formation are precisely this subclass, in which the carbon enrichment reflects the chemical composition of the cloud they were formed and no mass donation from a binary companion is observed. Since stars in the giant branch convert carbon to nitrogen, to ensure a fair comparison with S2 members, we have selected from the literature, when available, carbon abundances already corrected by this effect \citep{pla14b}. That fact makes easier to analyze absolute abundances instead of element ratios even for evolved stars where the mixing process are playing a remarkable roll.

\begin{figure}
	\includegraphics[width=57mm,angle=90]{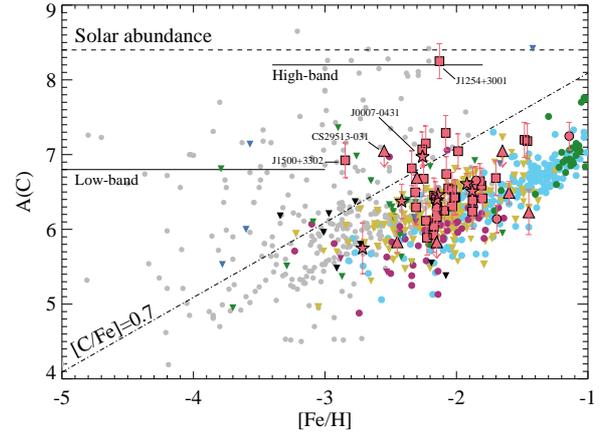}
        \caption{Carbon abundances A(C) of metal-poor stars as a function of [Fe/H] for S2 members. Symbols and references are the same as in Fig. \ref{chemistry}. The dashed line is the solar absolute carbon abundance, while the dotted-dashed line is the CEMP limit at [C/Fe]$=+0.7$. Thick lines are the approximate levels for low- and high-carbon bands, according to the groups qualitatively described in \citet{boni15} and \citet{yoo16}. S2 members (red circles, stars, squares and triangles according to the source of the data) are located in three different areas of the diagram.}
    \label{carbon}
\end{figure}

Section \ref{sec:analisys} presents 62 carbon measurements from S2 members and 63\% of them are VMP. Fig.~\ref{carbon} shows the absolute carbon abundance versus metallicity diagram. Some 20\% of the VMP stars are found to be CEMP stars. This ratio is in good agreement with the dwarf galaxies in the cosmological simulations of \citet{salv15}. In the case of J0007$-$0431, we have high-resolution analysis showing no n-capture process enrichment at all, so this star is classified as CEMP-no star. In the plot, bands are shown marking the regions of the A(C)-[Fe/H] diagram in which CEMP-no and the other n-enriched subclasses of CEMP stars reside. The S2 members are located in three different parts of the diagram:

\begin{itemize}
    \item Carbon-normal [C/Fe]$<0.7$ members: This group represents
      $\sim88\%$ of the S2 members and lies below the
      dotted-dashed line ([Fe/H]$<0.7$).
    \item CEMP-low members: With $\sim10\%$ of the stars belonging to this group, these are likely CEMP-no stars as J0007$-$0431 is.
    \item CEMP-high members: There is one star (J1252+3001) that shows almost the same carbon abundance as the Sun but more than a hundred times less metallic content. This could arise from mass transfer from an AGB binary companion. 
\end{itemize}

We also show in Fig. \ref{carbon} the carbon distribution function for other dwarf galaxies. With the exception of Segue 1 \citep{fresegue14}, most of them have members that are carbon-normal, like the majority of S2 members. However, S2 also has a significant fraction of CEMP-no stars. This suggests, similar to Segue 1, that S2 is indeed a very primitive galaxy with members formed from the pollution of a very few CCSNe. Since Segue 1 members are more metal-poor, we expect that S2 had a slightly faster chemical evolution corresponding to a higher mass system. Interestingly enough, both galaxies do contain at least one strongly carbon-enhanced star that lies in the high band of the diagram. 
The number of dwarf galaxy members in Fig. \ref{carbon} that
show carbon enhancement and no significant n-capture process enrichment is quite low (e.g., Figure 1 in \citet{yoon19}). However, according to \citet{salv15}, there should be CEMP-no stars in every dwarf galaxy so S2 supports this result.

\subsubsection{Aluminium and iron-peak elements}

In the upper-row of Fig. \ref{chemistry}, we show the Al behaviour of S2 members. This seems to agree with other metal-poor stars in the Milky Way. Low [Al/Fe] values are typical signatures of dwarf galaxies, in contrast with globular clusters (GC) \citep{cavallo96}. In addition, the three RGB stars (red circles and triangles) with Al measurement do not show significant Al enhancement but are compatible with the behaviour of dwarf S2 members. That fact confirms the idea that proton-capture process that converts Mg into Al is not reachable at RGB temperatures \citep{spite06}.

Sc and iron peak-elements such as Cr, and Co show compatible trends with other dwarf galaxies. The Co ratio is almost flat with metallicity, whereas Sculptor shows Co decreasing with higher metallicity. Finally, V is not very much studied in dwarf galaxies at the time of this writing, but its behaviour in S2 is similar to the halo field stars. However, Mn and Ni show a slightly different behaviour, with higher dispersion with metallicity and the largest overabundance in the iron-peak family. Theoretical studies on SN Ia yields by \citet{kobayashi20} show an anticorrelation of $\alpha$-elements with Mn and lower Ni production in dwarf galaxies. We do not see a Mn trend in Fig.~\ref{chemistry} but a clear Ni overabundances in two different sequences is shown. 

\subsubsection{Sodium and Nickel: a puzzling relation}\label{sec:nina}

Sodium, and also nickel production, are moderated by neutron excesses in supernovae \citep{tie90}. Only nickel is significantly produced by SN Ia \citep{suj95}. Sodium is manufactured in massive stars by helium burning and expelled by SN II. So, we expect the ratios [Ni/Fe] and [Na/Fe] to show some correlation, at least in the early stages of chemical evolution. When SN Ia increase the nickel production, the correlation is affected, at the same point when the [$\alpha$/Fe] and [Na/Fe] ratios start to decline. So, the combination of sodium and nickel has been used to compare different contributions from SN Ia and SN II in halo stars \citep{niss97, niss10}. They found a clear relation between [Na/Fe] and [Ni/Fe] in $\alpha$-deficient stars with high Galactocentric distances. They argued that the outer halo is mostly composed by accreted dSphs and the Ni$-$Na relation is a merger indicator. However, further studies by \citet{venn04} pointed out this relation is less clear in accreted dwarf galaxies than in halo stars. They suggested the relation is naturally produced in massive stars. In addition, at lower metallicities, the behaviour of Na in dSphs is halo-like, but not at higher metallicities ([Fe/H]$>-1$) \citep{To09}.  
 
\begin{figure}
	\includegraphics[width=57mm,angle=90]{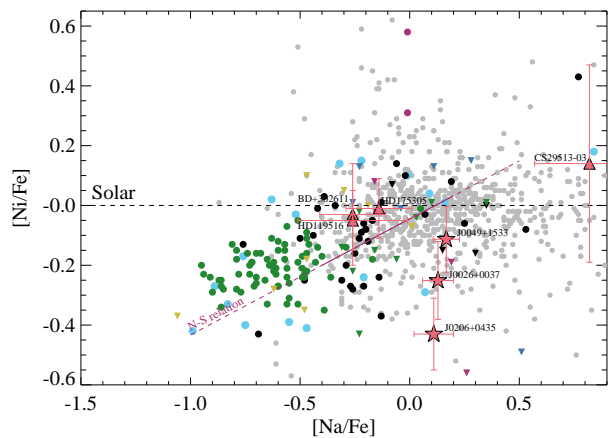}
        \caption{Sodium and nickel abundances relative to iron for
         S2 members. Symbols and
          references are the same as in Fig. \ref{chemistry}. The
          solid red line represents the [Ni/Fe] vs [Na/Fe] low found by \citet{niss97, niss10} for halo stars with
          $\alpha$-deficiency. The red dashed is the prolongation of this relation. The S2 members appear to show a very different behaviour.}
    \label{nina}
\end{figure}

In Fig.~\ref{nina}, we show the [Ni/Fe] versus [Na/Fe] relation for
seven S2 members (red stars and triangles), together with other stars from dwarf
galaxies. The original relation found by \citet{niss97} for
$\alpha$-deficient halo stars is represented by a pink thick line. Fornax
members (green circles) seem to roughly follow the Nissen-Schuster (N-S) relation.  
 However, the S2 stars seem to behave in two different ways: i) The main-sequence/SGB population has decreasing [Ni/Fe] at higher metallicities for almost the same [Na/Fe] ratio, with the exception of CS29513$-$03 that shows a similar trend but with higher [Na/Fe] ratio. This sodium enhancement is less likely to be attributed to mixing process in low giant branch phase \citep{gratton00}.  ii) The evolved RGB population is clearly sodium-poor and fits nicely within the N-S relation. While in globula rclusters, the O-Na anticorrelation is observed in the upper-RGB, there is no evidence of sodium enhancement in halo field stars \citep{gratton00}.
 
Since no further mixing episodes seem to be affecting S2 members, we suggest these two S2 groups (Na-rich and Na-poor) reflect the chemical signature of two different formation environments. One of them follows the N-S relation, whereas the other seems to behaves quite different.  This is consistent with the idea that the [Ni/Fe] versus [Na/Fe] relation is not a dSph indicator \citep{venn04}, and suggests that other mechanisms may well be playing a role.

\subsection{Comparison to Roederer et al. (2010)}

At the metal-poor end, the high-resolution abundances obtained as part
of this study agree well with those reported in
\citet{Roederer2010}. However, as Figures~\ref{chemistry},
\ref{alpha}, \ref{neutro} and \ref{nina} demonstrate, the three most
metal-rich stars retained from \citet{Roederer2010}, i.e. BD +30 2611,
CD --36 1052 and HD 175305 appear to follow trends distinct from the
rest of the sample. In Figure~\ref{chemistry}, these stars (filled in
triangles) appear peculiar in Sc as well as in Mg, Ca and Ti. In
$\alpha$-elements, instead of tracing the decline to Solar values,
rather oddly, these three stars tend to levitate at the plateau level,
the behaviour most noticeable in Ca. Furthermore in Ni (left panel in
the bottom row of Figure~\ref{chemistry}) two different trends are
noticeable: the metal-poor stars show a climb-down from the Solar
abundance ratio, yet at higher [Fe/H], BD +30 2611, CD--36 1052 and HD
175305 display values of [NiI/Fe] reset back up close to the Solar
value. Disagreement between the Ni abundances for these three stars
and our most recent measurements are highlighted in
Figure~\ref{nina}. In the plane of [Ni/Fe] and [Na/Fe], the filled
triangles \citep[][]{Roederer2010} occupy an entirely different region
compared to filled stars (this study). Finally, as Figure~\ref{neutro}
demonstrates, the three metal-rich stars from \citet{Roederer2010}
attain much higher [Sr/Fe] ratios compared to the rest of our
sample. Interestingly, a look at the top panel of Figure~\ref{param}
reveals that these three stars appear to be the most evolved in our
combined sample: their location is consistent with AGB/HB membership. Finally, we remark rotation \citep{behr00} and turbulence \citep{michaud08} 
effects could potentially lead to elemental abundance anomalies in this
kind of evolved stars.

\section{Conclusions}\label{sec:conclusions}

We have used {\it Gaia} DR2 astrometry to refine the kinematic
identification of the nearby halo stream S2, building a catalogue of 60
high-confidence members originally observed by SDSS, LAMOST and
APOGEE. We have re-analyzed the archival low-resolution spectroscopy
complemented by high S/N observations with HET/LRS2 to derive stellar
parameters and carbon abundances consistently. Taking advantage of the
stream's proximity to the Sun, we have acquired high-resolution
spectroscopic data for 7 Main Sequence S2 member stars with VLT/UVES,
VLT/X-SHOOTER and GTC/HORuS and have measured the abundances of C, Li,
Na, Al, Sc, $\alpha$-elements (Mg, Si, Ca Ti), iron-peak elements (V,
Cr, Mn, Fe, Co, Ni), and n-capture process elements (Sr, Ba). Finally, our
high-resolution chemical abundances for high-surface gravity S2
members are combined with a revised and reduced sample of the likely
S2 giant stars obtained by \citet{Roederer2010}. We summarize the main
results of our analysis below.

\begin{itemize}

\item Based on the combination of low- and high-resolution abundance
  measurements, we conclude that the MDF of the
  S2 member stars spans the range of
  $-2.7<$[Fe/H]$<-1.5$. Additionally, there are two low-resolution
  spectroscopic S2 members with [Fe/H]$=-3$ and [Fe/H]=$-1.2$ respectively,
  possibly broadening the S2's MDF even further. Based on the revised
  orbital properties, we conclude that the two most metal-poor stars
  from the original study of \citet{Roederer2010} with [Fe/H]=$-3.3$
  and [Fe/H]=$-3.4$ cannot be assigned to S2 with
  confidence. Notwithstanding the uncertainty in the determination of
  the highest and the lowest levels of metal enrichment reached by the
  S2 stars, its MDF spans more than 1 dex in [Fe/H] and therefore we
  conclude that without a shred of doubt the stream's progenitor was a
  now defunct dwarf galaxy.
  
\item The average of the three $\alpha$-elements shows hints of a
  pattern consistent with a ``knee'' at [Fe/H]$<-2$, whose presence
  can not be established unambiguously. We note, however, that the
  individual tracks of Mg, Si, Ca and Ti as a function of [Fe/H]
  appear rather flat, yet the existence of a gentle downward slope
  with increasing metallicity can not be ruled out. Importantly,
  throughout the entire [Fe/H] range probed by the S2 members, the [Ti/Fe]
  ratio stays firmly above that measured for the stars in the Sculptor
  dwarf satellite. In light of the behaviour of the S2 $\alpha$
  abundances, we hypothesise that the progenitor galaxy was barely
  affected by the SN Ia pollution. S2's star formation proceeded
  slowly and was cut short by its accretion and disruption by the
  Milky Way.

\item The behaviour of the neutron-capture elements in S2 is
  remarkable. On one hand, the [Ba/Fe] ratio is at the level consistent
  with the bulk of the halo stars and that of the classical dwarfs,
  like Sculptor and the globular clusters. This can be contrasted with
  the Ba ratios for the Galactic ultra-faint dwarfs, for which all but one
  (Reticulum 2) show an under-abundance of Ba compared to the field. On the
  other hand, many S2 stars with [Fe/H]$<-2$ show depleted levels of
  Sr, clearly below the halo trend at these metallicities and fully
  consistent with the behaviour of most of the Milky Way's
  UFDs. Finally, the ratios of Sr and Ba (especially at [Fe/H]$<-2$)
  show an increased scatter compared to both the halo and dwarfs like
  Sculptor, indicative of the stochasticity of the enrichment
  process. The Sr/Ba pattern exhibited by the S2 stream is unique amongst the
  studied Galactic subsystems and points to the existence of multiple
  (additional) sites of the production of neutron-capture elements.

\item Our study provides new insights into the problem of the Na-Ni
  relation with 3 measurements that do not seem to agree with the
  previously established Nissen-Schuster relation for slightly more
  metal-rich halo population.
  
  \item Finally, we also find
  S2 members are displayed in different regions of the $\rm A(C)-[Fe/H]$ diagram
  with a significant fraction,  $\sim10$\%,  located in the ancient CEMP-no group.
  This old population, rarely observed in other dwarf galaxies, supports the conclusion
  S2 is the shreds of a very unevolved progenitor system.
\end{itemize}

Taken together, the enrichment patterns of the elements available as
part of our study of the S2 stream members point to a progenitor
galaxy which is unlike any other dwarf satellite on orbit around the
Milky Way. Star-formation in the S2 dwarf appears to have reached
significant metallicity levels of [Fe/H]$\sim-1.5$ or perhaps even
[Fe/H]$\sim-1.2$, yet with little evidence for SN Ia contribution. This
primitive enrichment pattern is at odds with the absence of extremely
metal poor stars in S2 and could instead indicate possible pollution
from the nearby large galaxy, e.g. Milky Way. S2's progenitor passes
many chemical tests for a building block of the Galactic stellar
halo, but fails spectacularly at the last hurdle - the neutron-capture
element ratio. Our study provides a unique and previously unexplored
view of star-formation and chemical enrichment at the high
redshift, but it still leaves some puzzling questions unanswered.

\section{Data Availability and online material}
All the data reduced and analyzed for the present article is fully available under request to the corresponding author\footnote{daguado@ast.cam.ac.uk}. Also, the UVES and X-SHOOTER data used in this article can be found in the original ESO archive\footnote{ https://archive.eso.org/}. We also include as online material a summary of the line list with those detected in S2.

\section*{Acknowledgements}
We thank Vanessa Hill for very useful comments about the nature of
Sculptor galaxy. Kim Venn very kindly provided interesting feedback about Carina galaxy. We also thank Chris Tout for highly appreciated suggestions, and the anonymous referee for a useful report.
This work is partly based on data from the GTC Public
Archive at CAB (INTA-CSIC), developed in the framework of the Spanish
Virtual Observatory project supported by the Spanish MINECO through
grants AYA 2011-24052 and AYA 2014-55216. The system is maintained by
the Data Archive Unit of the CAB (INTA-CSIC). We would like to thank
GRANTECAN staff members for their efficiency during the observing runs
and all the effort done during HORuS commissioning.The Low Resolution 
Spectrograph 2 (LRS2) was developed and funded by the University of Texas
at Austin McDonald Observatory and Department of Astronomy and by Pennsylvania
State University. We thank the Leibniz-Institut für Astrophysik Potsdam (AIP) and 
the Institut für Astrophysik Göttingen (IAG) for their contributions to the construction 
of the integral field units.  Some of the figures within
this paper were produced using IDL colour-blind-friendly colour tables
\citep[see][]{pjwright17}.  DA thanks the Leverhulme Trust for
financial support.
SK is partially supported by NSF grants AST-1813881, AST-1909584 and 
Heising-Simons foundation grant 2018-1030.

C.A.P. and J.I.G.H. acknowledge financial support from the Spanish Ministry of Science and Innovation (MICINN) project AYA2017-86389-P. J.I.G.H. also acknowledges financial support from Spanish MICINN under the 2013 Ram\'on y Cajal program RYC-2013-14875.
G. A. Lanfranchi thanks the Brazilian agency FAPESP (grants 2014/11156-4 and 2017/25799-2).




\bibliographystyle{mnras}

\bibliography{biblio}


\begin{appendix}

\section{Lithium: a fossil record from early Universe}\label{sec:lithium}

Primordial lithium from Big Bang Nucleosythesis (BBN) could be in
principle measured in the atmospheres of unevolved metal-poor
stars. \citet{spi82} and \citet{reb88} noticed an almost constant
value for lithium abundance from the analysis of tens of VMP stars,
A(Li)=$2.2\pm0.1$. This value is named Lithium plateau or Spite
plateau. Since then the regime on which the plateau has been observed
has been expanded down to a metallicity [Fe/H]$<-6.1$
\citep{agu19b}. High resolution observations have been also detected the plateau in $\omega$~Centauri
\citep{mona10}, the globular cluster M54 \citep{mucc14} and Sculptor
dwarf galaxy \citep{hill19}. More recently, \citet{mola20} also found
the signature of the Lithium plateau in the major merger event in the {\it Gaia}-Sausage
\citep{belo18, mye18c}. That fact suggests whatever thing is
responsible for the existence of the plateau is also happening outside
of the Milky Way. Intriguingly, thanks to cosmological analysis of the
Cosmic Microwave Background (CMB) the primordial lithium is calculated
to be 2-3 times larger than observed in old stars
\citep{spe03,coc13}. This discrepancy is known as the cosmological
lithium problem and is a longstanding problem in modern cosmology
~\citep[see e.g.,][and references therein]{fie11,fie20}.

\begin{figure}
	\includegraphics[width=57mm,angle=90]{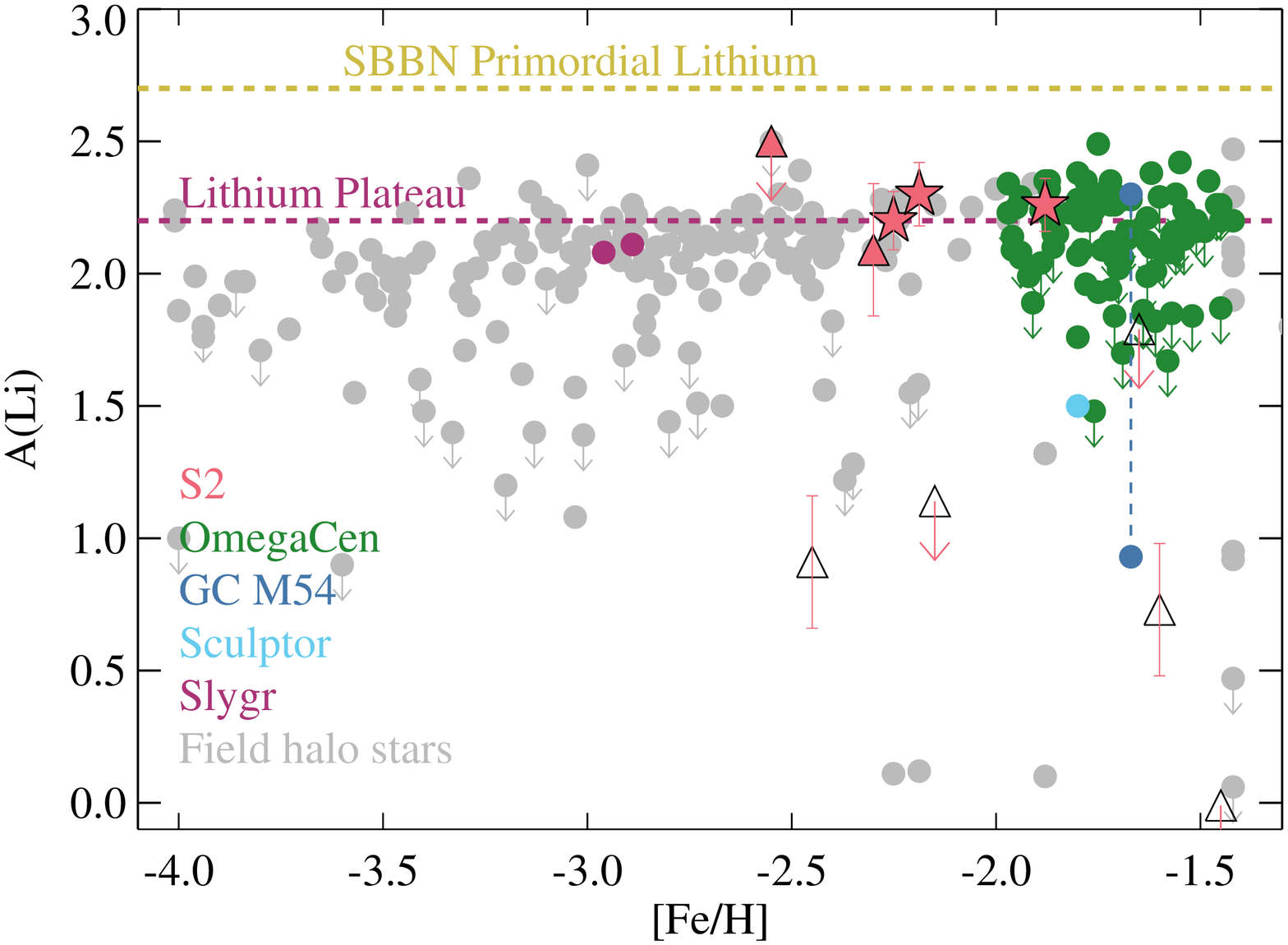}
        \caption{ Lithium abundance, A(Li), versus metallicity,
          $\left[{\rm Fe/H}\right]$, of the S2 stars observed with
          UVES (red stars) and MIKE (red triangles) compared with the
          values from $\omega$~Centauri \citep{mona10}, the globular
          cluster M54 \citep{mucc14}, the Sculptor galaxy
          \citep{hill19}, the Slygr stream \citep{roe19}, and other
          dwarf - turn-off stars ($\log g \geq 3.7$) with Li abundance
          values from
          \citet{asp06II,jon08b,aoki09,mel10,sbo10,aoki12,sbo12,mas12,boni12,han14,han15I,mat17,boni18,agu19a,jon19}
          and references therein. We also include evolved S2 members
          from \citet{Roederer2010} (uncolored triangles) in which
          lithium is already depleted. The Lithium plateau (also
          called “Spite Plateau”) reference is shown as solid line at
          a level of A(Li)~$=2.20$\,dex. Blue dashed line represents
          the primordial lithium value (A(Li)$\sim$2.7) from WMAP
          \citep{spe03,coc13}}
    \label{lithium}
\end{figure}

We are able to measure the lithium double at 6707.8\AA\, in three S2
members, J0026+0037, J0049+1533, and J0206+0435. The values we derive
are fully compatible with the plateau, A(Li)=$2.3\pm 0.1$, $2.20\pm
0.1$, and $2.26\pm 0.1$ respectively, exacerbating the lithium
problem. The level of lithium measured in these stars is another
indicator of the slow chemical evolution suffered by S2. In Figure
\ref{lithium} we show the lithium measurements for unevolved
metal-poor stars in S2 (red symbols), $\omega$~Centauri \citep{mona10}
(green circles), the globular cluster M54 \citep{mucc14} (dark blue
circle), the
Slygr stream \citep{roe19} (purple circles) and more Milky Way
main-sequence turn-off halo stars from literature (grey circles). 
For completeness we add uncoloured triangles that represent more evolved S2
stars and a light blue circle that represents a measurement from a giant star in Sculptor galaxy \citep{hill19}. As expected we clearly see how after dredge-up and mixing
processes lithium is diluted \citep{brown89, gratton00}. Fig. \ref{lithium}
clearly suggest the lithium behaviour we find in canonical MW halo
stars is the same than accreted satellites and extragalactic
structures such as $\omega$~Centauri. We argue that the lithium
abundance of the oldest stars is not affected by the environment they
were formed and suggest the cosmological lithium problem cannot be
exclusively explained by near cosmology effects.

\section{Tables}
\begin{table*}
\begin{center}
\renewcommand{\tabcolsep}{5pt}
\centering
\caption{Coordinates and stellar parameters from low-resolution analysis.
\label{ape1}}
\begin{tabular}{llcccccccccc}
\hline
S2 member& & $V$ &      RA &      DEC &$T_{\rm eff}$ & $\log g$  & $\left[{\rm Fe/H}\right]$
& $\left[{\rm C/Fe}\right]$ &$\rm <S/N>$& Flag\\
    & & mag & h  '   ''&$\mathring{}$  '  ''   & K & $\rm cm\, s^{-2}$ &               &             &          \\
\hline
SDSS  &J0004$-$0442&       17.4& 00:04:37.15& $-$04:42:32.8&       5190$\pm$      101&       4.65$\pm$     0.50&      -1.87$\pm$     0.10&     -0.28$\pm$     0.20&       51& \\
SDSS  &J0007$-$0431&       17.5& 00:07:05.36& $-$04:31:47.7&       6264$\pm$      103&       3.62$\pm$     0.53&      -2.10$\pm$     0.10&      0.71$\pm$     0.27&       55&1,4 \\
SDSS  &J0009$-$0548&       18.5& 00:09:30.31& $-$05:48:57.8&       5955$\pm$      103&       4.99$\pm$     0.51&      -2.31$\pm$     0.11&      0.40$\pm$     0.21&       42& \\
SDSS  &J0026+0037&       15.8& 00:26:19.82&  00:37:34.7&       6305$\pm$      102&       4.04$\pm$     0.52&      -2.02$\pm$     0.10&      0.70$\pm$     0.24&       81&1,4 \\
SDSS  &J0028$-$0932&       17.7& 00:28:25.11& $-$09:32:29.2&       6269$\pm$      110&       4.88$\pm$     0.52&      -2.34$\pm$     0.11&      0.75$\pm$     0.22&       57& \\
SDSS  &J0032+0748&       17.6& 00:32:21.81&  07:48:05.1&       5963$\pm$      101&       4.35$\pm$     0.50&      -1.82$\pm$     0.10&      0.06$\pm$     0.20&       71& \\
SDSS  &J0033$-$0018&       18.2& 00:33:58.09& $-$00:18:51.6&       5998$\pm$      104&       4.80$\pm$     0.51&      -2.24$\pm$     0.11&      0.52$\pm$     0.21&       32&3 \\
SDSS  &J0033$-$0928&       19.4& 00:33:58.97& $-$09:28:18.0&       5951$\pm$      100&       4.99$\pm$     0.50&      -1.63$\pm$     0.10&      0.31$\pm$     0.20&       14& \\
SDSS  &J0041+0723&       17.9& 00:41:17.93&  07:23:41.9&       6267$\pm$      102&       3.77$\pm$     0.52&      -1.87$\pm$     0.10&     -0.13$\pm$     0.39&       31& \\
SDSS  &J0045+1345&       17.4& 00:45:00.91&  13:45:46.0&       5814$\pm$      101&       4.51$\pm$     0.50&      -2.16$\pm$     0.10&      0.22$\pm$     0.20&       75& \\
SDSS  &J0047+1433&       18.2& 00:47:55.29&  14:33:55.6&       5434$\pm$      101&       4.82$\pm$     0.50&      -1.80$\pm$     0.10&     -0.18$\pm$     0.20&       52& \\
SDSS  &J0049+1533&       15.6& 00:49:39.99&  15:33:17.5&       6315$\pm$      102&       4.06$\pm$     0.52&      -1.94$\pm$     0.11&      0.14$\pm$     0.63&      104&1 \\
SDSS  &J0102$-$0043&       17.8& 01:02:00.84& $-$00:43:16.2&       5341$\pm$      101&       4.54$\pm$     0.50&      -2.14$\pm$     0.10&      0.05$\pm$     0.20&       49& \\
SDSS  &J0140$-$0826&       18.5& 01:40:35.69& $-$08:26:14.7&       6185$\pm$      101&       4.41$\pm$     0.50&      -1.48$\pm$     0.10&      0.28$\pm$     0.20&       46& \\
SDSS  &J0148+0018&       18.5& 01:48:58.66&  00:18:51.4&       4844$\pm$      101&       5.00$\pm$     0.50&      -2.09$\pm$     0.10&     -0.05$\pm$     0.20&       37& \\
LAMOST&J0206+0435&       15.8& 02:06:53.72&  04:35:44.5&       6143$\pm$      105&       4.63$\pm$     0.51&      -1.99$\pm$     0.11&      0.29$\pm$     0.22&      102&1 \\
SDSS  &J0224$-$0012&       17.9& 02:24:38.61& $-$00:12:40.8&       5024$\pm$      101&       4.91$\pm$     0.50&      -2.17$\pm$     0.10&     -0.19$\pm$     0.20&       56& \\
SDSS  &J0239+2615&       17.2& 02:39:29.14&  26:15:00.2&       5350$\pm$      101&       4.70$\pm$     0.50&      -2.03$\pm$     0.10&     -0.06$\pm$     0.20&      105& \\
SDSS  &J0330+0009&       17.7& 03:30:52.99&  00:09:16.6&       5010$\pm$      101&       4.99$\pm$     0.50&      -2.14$\pm$     0.10&     -0.08$\pm$     0.20&       47& \\
SDSS  &J0445+0041&       17.2& 04:45:01.46&  00:41:28.2&       5647$\pm$      101&       4.89$\pm$     0.50&      -2.15$\pm$     0.10&      0.16$\pm$     0.20&       77& \\
LAMOST&J0808+2418&       15.6& 08:08:33.43&  24:18:46.8&       6070$\pm$      102&       4.50$\pm$     0.51&      -2.38$\pm$     0.10&      0.20$\pm$     0.35&       71&1,4 \\
SDSS  &J0813+3658&       19.0& 08:13:32.56&  36:58:05.8&       6319$\pm$      112&       4.99$\pm$     0.52&      -2.99$\pm$     0.44&     -0.49$\pm$     4.94&       20&2,4 \\
SDSS  &J0829+3126&       18.9& 08:29:07.76&  31:26:20.7&       5874$\pm$      104&       4.98$\pm$     0.52&      -2.90$\pm$     0.10&      0.67$\pm$     0.21&       11&2,3,4 \\
LAMOST&J0853+2522&       17.6& 08:53:01.53&  25:22:14.5&       5191$\pm$      101&       4.53$\pm$     0.50&      -2.22$\pm$     0.10&     -0.28$\pm$     0.20&       51& \\
LAMOST&J0917+2319&       16.3& 09:17:59.68&  23:19:01.9&       6179$\pm$      104&       3.16$\pm$     0.57&      -2.27$\pm$     0.37&     -0.56$\pm$     2.00&       21&4 \\
SDSS  &J0929+4105&       15.6& 09:29:40.68&  41:05:52.2&       5973$\pm$      105&       4.64$\pm$     0.51&      -2.19$\pm$     0.11&      0.36$\pm$     0.21&      100&1 \\
SDSS  &J0933+2941&       18.8& 09:33:01.73&  29:41:08.7&       4982$\pm$      101&       4.99$\pm$     0.50&      -2.17$\pm$     0.10&     -0.29$\pm$     0.20&       37& \\
SDSS  &J0949+5013&       18.6& 09:49:12.95&  50:13:39.6&       5764$\pm$      100&       3.56$\pm$     0.50&      -1.66$\pm$     0.10&      0.24$\pm$     0.20&       26&2,3 \\
SDSS  &J1017+3647&       18.5& 10:17:31.15&  36:47:51.9&       6206$\pm$      101&       4.47$\pm$     0.50&      -1.74$\pm$     0.10&      0.20$\pm$     0.22&       31& \\
SDSS  &J1053+0349&       17.8& 10:53:22.47&  03:49:44.8&       5369$\pm$      101&       4.43$\pm$     0.50&      -2.23$\pm$     0.10&     -0.05$\pm$     0.20&       48& \\
SDSS  &J1103+4956&       18.2& 11:03:13.46&  49:56:44.3&       5849$\pm$      101&       4.10$\pm$     0.50&      -2.07$\pm$     0.10&      0.41$\pm$     0.20&       25&2 \\
SDSS  &J1122+4605&       19.2& 11:22:16.83&  46:05:56.3&       6007$\pm$      100&       4.55$\pm$     0.50&      -1.46$\pm$     0.10&      0.24$\pm$     0.20&       23& \\
SDSS  &J1124+0202&       17.0& 11:24:26.02&  02:02:05.5&       6177$\pm$      103&       4.07$\pm$     0.52&      -2.30$\pm$     0.11&      0.20$\pm$     0.89&       51& \\
SDSS  &J1129+1005&       18.4& 11:29:31.74&  10:05:00.9&       6203$\pm$      103&       4.13$\pm$     0.52&      -2.26$\pm$     0.10&      0.93$\pm$     0.21&       34& \\
SDSS  &J1132+2856&       16.4& 11:32:23.00&  28:56:23.3&       5946$\pm$      101&       4.19$\pm$     0.50&      -2.00$\pm$     0.10&      0.03$\pm$     0.21&       65& \\
SDSS  &J1159+2511&       18.3& 11:59:53.78&  25:11:45.3&       6444$\pm$      109&       4.90$\pm$     0.51&      -1.98$\pm$     0.12&      0.63$\pm$     0.24&       23& \\
SDSS  &J1204+2110&       19.6& 12:04:36.13&  21:10:38.4&       5737$\pm$      101&       4.99$\pm$     0.50&      -1.70$\pm$     0.10&     -0.01$\pm$     0.20&       19& \\
SDSS  &J1207+1926&       17.5& 12:07:44.94&  19:26:54.2&       6235$\pm$      102&       4.45$\pm$     0.51&      -1.92$\pm$     0.10&      0.01$\pm$     0.26&       55&4 \\
LAMOST&J1243+1003&       16.6& 12:43:22.31&  10:03:47.0&       6313$\pm$      107&       4.73$\pm$     0.51&      -1.81$\pm$     0.12&     -0.00$\pm$     0.25&       32& \\
SDSS  &J1244$-$0200&       16.5& 12:44:21.07& $-$02:00:08.5&       6314$\pm$      102&       4.48$\pm$     0.51&      -2.06$\pm$     0.10&      0.21$\pm$     0.57&      111& \\
SDSS  &J1253+3001&       16.8& 12:53:16.92&  30:01:38.8&       6092$\pm$      102&       3.25$\pm$     0.51&      -2.12$\pm$     0.10&      1.97$\pm$     0.20&       58& \\
SDSS  &J1309+1847&       19.0& 13:09:43.14&  18:47:54.0&       5684$\pm$      100&       4.99$\pm$     0.50&      -2.08$\pm$     0.10&      0.97$\pm$     0.20&       20& \\
SDSS  &J1327+2232&       19.0& 13:27:54.58&  22:32:41.1&       5962$\pm$      101&       1.93$\pm$     0.54&      -2.38$\pm$     0.11&      0.94$\pm$     0.26&       20&4 \\
SDSS  &J1417+3514&       18.4& 14:17:55.92&  35:14:03.8&       6069$\pm$      102&       3.90$\pm$     0.51&      -2.07$\pm$     0.10&     -0.01$\pm$     0.26&       27&4 \\
SDSS  &J1451+3313&       18.2& 14:51:22.53&  33:13:59.7&       5408$\pm$      101&       4.18$\pm$     0.50&      -2.01$\pm$     0.10&      0.03$\pm$     0.20&       40& \\
SDSS  &J1500+3302&       18.2& 15:00:14.44&  33:02:13.9&       6214$\pm$      104&       3.88$\pm$     0.55&      -2.84$\pm$     0.10&      1.37$\pm$     0.24&       40&2 \\
SDSS  &J1554+4450&       18.5& 15:54:00.91&  44:50:14.2&       6125$\pm$      102&       4.25$\pm$     0.51&      -2.03$\pm$     0.10&      0.16$\pm$     0.26&       26& \\
SDSS  &J1703+2317&       17.6& 17:03:35.08&  23:17:51.5&       5251$\pm$      101&       4.53$\pm$     0.50&      -2.07$\pm$     0.10&     -0.11$\pm$     0.20&       56&2,4 \\
SDSS  &J2329$-$1009&       17.4& 23:29:56.24& $-$10:09:34.3&       5705$\pm$      101&       4.45$\pm$     0.50&      -1.87$\pm$     0.10&      0.07$\pm$     0.20&       79& \\
SDSS  &J2340+0046&       18.3& 23:40:56.86&  00:46:38.3&       6178$\pm$      102&       3.75$\pm$     0.52&      -2.06$\pm$     0.10&      0.75$\pm$     0.22&       42& \\
SDSS  &J2345$-$0003&       17.9& 23:45:52.73& $-$00:03:05.0&       6380$\pm$      103&       4.16$\pm$     0.53&      -2.31$\pm$     0.10&      0.75$\pm$     0.27&       65&1 \\
SDSS  &J2355+0015&       17.8& 23:55:04.76&  00:15:34.2&       6464$\pm$      120&       4.70$\pm$     0.52&      -2.23$\pm$     0.12&      0.98$\pm$     0.22&       34&4 \\
\hline
\multicolumn{5}{l}{1: Objects with high-resolution analysis.} \\
\multicolumn{5}{l}{2: Objects with positive $\rm v_{z}$ motion} \\
\multicolumn{5}{l}{3: Objects with tentative [Fe/H] and [C/Fe] from low-resolution analysis} \\
\multicolumn{5}{l}{4: Objects with tentative [C/Fe] from low-resolution analysis} \\
\hline
\end{tabular}
\end{center}
\end{table*}

\begin{table*}
\begin{center}
\renewcommand{\tabcolsep}{5pt}
\centering
\caption{APOGEE abundances for S2 members from ASPCAP.
\label{ape2}}
\begin{tabular}{lccccccccccc}
\hline
APOGEE name&$\rm T_{eff}$ & log $g$ &[Fe/H]&[C/Fe]&[Mg/Fe]&[Al/Fe]&[Si/Fe]&[Ca/Fe]&[Mn/Fe]&[Ni/Fe]&FLAG\\
\hline         
2M02452900-0100541&       4168$\pm$      15&       0.7$\pm$0.1&      -1.7$\pm$    0.1&     -0.6$\pm$    0.1&          0.2$\pm$    0.1&     -0.4$\pm$    0.1&      0.2$\pm$    0.1&      0.0$\pm$    0.1&           --&          0.0$\pm$    0.1&\\
2M11503654+5407268&       4907$\pm$      25&       2.1$\pm$0.1&      -1.1$\pm$    0.1&      0.0$\pm$    0.1&          0.1$\pm$    0.1&     -0.2$\pm$    0.1&      0.1$\pm$    0.1&      0.2$\pm$    0.1&           -0.4$\pm$    0.1&        -0.2$\pm$    0.1&\\
2M12250956-0057392&       5001$\pm$      26&       2.3$\pm$0.1&      -1.3$\pm$    0.1&      0.0$\pm$    0.1&          0.2$\pm$    0.1&     -0.3$\pm$    0.1&      0.2$\pm$    0.1&      0.2$\pm$    0.1&           -0.7$\pm$    0.1&          0.1$\pm$    0.1&1\\
2M13491436+2637457&       3456$\pm$      23&       4.7$\pm$0.1&      -0.5$\pm$    0.1&      0.0$\pm$    0.1&        -0.1$\pm$    0.1&     -0.1$\pm$    0.1&      0.0$\pm$    0.1&      0.2$\pm$    0.1&            0.1$\pm$    0.1&        -0.2 $\pm$    0.1&2\\
2M18470646+7443316&       5240$\pm$      46&       2.8$\pm$0.1&      -1.3$\pm$    0.1&     -0.1$\pm$    0.1&          0.1$\pm$    0.1&     -0.2$\pm$    0.2&      0.1$\pm$    0.1&      0.4$\pm$    0.1&              --             &            --           &1   \\
2M21312112+1307399&       4795$\pm$      16&       1.8$\pm$0.1&      -1.8$\pm$    0.1&      0.1$\pm$    0.1&          0.2$\pm$    0.1&     -0.4$\pm$    0.1&      0.2$\pm$    0.1&      0.1$\pm$    0.1&           -0.5$\pm$    0.1&          0.1$\pm$    0.1&\\
2M08422740+1303316&       5207$\pm$      16&       2.9$\pm$0.1&      -1.4$\pm$    0.1&      -0.2$\pm$    0.1&          0.2$\pm$    0.1&     -0.3$\pm$    0.1&      0.2$\pm$    0.1&      0.4$\pm$    0.1&           -0.1$\pm$    0.1&          0.2$\pm$    0.1&1\\
\hline
\multicolumn{5}{l}{1: Objects discarded by {\it Gaia} DR2 kinematics.} \\
\multicolumn{5}{l}{2: Objects flagged by ASPCAP and not used in this work.} \\

\hline
\end{tabular}
\end{center}
\end{table*}

\end{appendix}

\bsp	
\label{lastpage}
\end{document}